\documentclass[prd, superscriptaddress, tightenlines, longbibliography, nofootinbib, eqsecnum, amsfonts, amsmath, floatfix, twocolumn, notitlepage]{revtex4-2}
\pdfoutput=1

\usepackage[utf8]{inputenc}
\usepackage{mathrsfs}
\usepackage{euscript}
\usepackage{graphics}
\usepackage{graphicx}
\usepackage{amsmath}
\usepackage{amssymb}
\usepackage{gensymb}
\usepackage{bm}
\usepackage[usenames,dvipsnames,svgnames,table]{xcolor}
\definecolor{linkcolor}{rgb}{0.6,0,0}
\definecolor{citecolor}{rgb}{0,0,0.75}
\definecolor{urlcolor}{rgb}{0.12,0.46,0.7}
\usepackage{xspace}
\usepackage{wasysym}
\usepackage{times}
\usepackage{appendix}
\usepackage{comment}
\usepackage{lipsum}
\usepackage[nolist,nohyperlinks]{acronym}
\usepackage{float}
\usepackage{simplewick}
\usepackage{natbib, ifthen}
\usepackage[breaklinks, colorlinks, urlcolor=urlcolor, linkcolor=linkcolor,citecolor=citecolor,pdfencoding=auto]{hyperref}
\usepackage{longtable}
\usepackage{listings}

\newcommand{\fsky}{f_{\rm sky}}

\providecommand{\planck}{\textit{Planck}}
\providecommand{\Planck}{\planck}
\newcommand{\camspec}{{\tt CamSpec}}
\newcommand{\plik}{{\tt Plik}}

\newcommand{\mksym}[1]{\ifmmode {\rm #1}\else #1\fi}

\providecommand{\Planck}{\textit{Planck}}
\providecommand{\planck}{\Planck}

\providecommand{\gea}{\ga}

\providecommand{\agt}{\gea}
\providecommand{\text}[1]{\rm{#1}}

\renewcommand{\d}{\text{d}}

\providecommand{\CAMB}{{\tt camb}}

\providecommand{\Cobaya}{{\tt Cobaya}}

\providecommand{\LCDM}{{$\rm{\Lambda CDM}$}}

\providecommand{\HEALpix}{{\tt HEALpix}}

\newcommand{\begm}{\begin{pmatrix}}
\newcommand{\enm}{\end{pmatrix}}

\newcommand\ba{\begin{eqnarray}}
\newcommand\ea{\end{eqnarray}}
\newcommand\bea{\begin{eqnarray}}
\newcommand\eea{\end{eqnarray}}

\newcommand\be{\begin{equation}}
\newcommand\ee{\end{equation}}

\newcommand{\boldvec}[1]{{\mbox{\boldmath{$#1$}}}}

\newcommand{\vn}{\boldvec{n}}

\newcommand{\vnhat}{\hat{\vn}}

\defcitealias{PL2018}{PL2018}

\defcitealias{Carron:2022eyg}{PL4}

\newcommand{\JC}[1]{{#1}}

\newcommand{\Cov}[0]{ {\rm{Cov}} }

\newcommand{\ellsky}{\ell_{\rm sky}}
\newcommand{\msky}{m_{\rm sky}}

\newcommand{\av}[1]{\left \langle #1\right\rangle}

\newcommand{\Tfiras}{T_{\text{\sc FIRAS}}}
\newcommand{\Rtt}[0]{ {R^{TT}} }
\newcommand{\Rpt}[0]{ {R^{\phi T}} }
\newcommand{\TWF}[0]{T^{\rm WF}}
\newcommand{\TWFd}[0]{T^{\rm WF, \dagger}}

\begin{document}
\title{Planck ISW-lensing likelihood and the CMB temperature }

\newcommand{\Sussex}{Department of Physics \& Astronomy, University of Sussex, Brighton BN1 9QH, UK}
\newcommand{\Geneve}{Universit\'e de Gen\`eve, D\'epartement de Physique Th\'eorique et CAP, 24 Quai Ansermet, CH-1211 Gen\`eve 4, Switzerland}
\newcommand{\Cardiff}{School of Physics and Astronomy, Cardiff University, The Parade, Cardiff, CF24 3AA, UK}
\newcommand{\CCA}{Center for Computational Astrophysics, Flatiron Institute, 162 5th Avenue, 10010, New York, NY, USA}

\author{Julien Carron}
\affiliation{\Geneve}
\affiliation{\Sussex}
\author{Antony Lewis}
\affiliation{\Sussex}
\author{Giulio Fabbian}
\affiliation{\CCA}
\affiliation{\Cardiff}
\affiliation{\Sussex}

  \begin{abstract}

  We present a new \planck\ CMB lensing-CMB temperature cross-correlation likelihood that can be used to constrain cosmology via the Integrated Sachs-Wolfe (ISW) effect.
  CMB lensing is an excellent tracer of ISW, and we use the latest PR4 \planck\ data maps and lensing reconstruction to produce the first public \planck\ likelihood to constrain this signal.
  We demonstrate the likelihood by constraining the CMB background temperature from \planck\ data alone, where the ISW-lensing cross-correlation is a powerful way to break the geometric degeneracy, substantially improving constraints from the CMB and lensing power spectra alone.
  \end{abstract}


   \maketitle

\section{Introduction}

The integrated Sachs-Wolfe effect (ISW, ~\citep{Sachs:67}) describes how photons pick up a net blue or redshift while propagating through time-varying potentials between last scattering and when we observe them today. In terms of the Weyl potential $\Psi$, ISW imprints a temperature perturbation
\begin{equation}\label{ISWint}
  \Delta T(\vnhat)_{\rm ISW} \approx 2\int_0^{\chi_*} \d \chi \dot\Psi(\chi\vnhat, \eta_0 - \chi),
\end{equation}
where a dot denotes conformal time derivative, $\eta_0$ is the conformal time today, and the integral is along the line of sight in direction $\vnhat$ between us and last scattering at comoving distance $\chi_*$. In a standard cold matter-dominated universe, linear gravitational potentials are constant because there is an exact compensation between decay due to expansion (the separation between comoving masses gets larger), and growth of the density perturbations (density perturbations grow proportional to the scale factor during matter domination). In the late universe, dark energy relatively increases the expansion rate, leading to a net decay in the amplitude of potentials with time, and hence a net ISW effect. The ISW is therefore a probe of the late-time density perturbations, with amplitude that depends on the dark-energy evolution~\cite{Crittenden:1995ak}, any modification of gravity \cite[e.g.][]{Kable:2021yws}, or other beyond flat-$\Lambda$CDM perturbation growth (for example curvature, dark matter interactions, massive neutrinos, etc.~\cite[e.g.][]{Lesgourgues:2007ix}).

The CMB lensing potential is correlated to ISW because the same gravitational potentials cause both effects. This is dominated by the late-time ISW signal from the dark energy era, which has significant contributions to distances about 1/3 of the way to last scattering. The early-ISW signal from potentials near recombination (due to the radiation density) is not significantly correlated to the lensing signal because it is produced very close to the last-scattering surface. The lensing potential-ISW correlation is therefore a probe of dark energy.

Unfortunately, the ISW signal cannot be measured independently as we only have access to the total temperature anisotropies including the sources from recombination. In practice, the primordial fluctuations dominate in most cosmologies, so that their cosmic variance acts as an irreducible source of noise for the temperature-lensing cross-correlation signal. In principle, this can be improved slightly by also using polarization to constrain the primordial anisotropies, but even with perfect observations the total signal remains relatively low. This is because the signal is limited to large scales: for small-scale perturbations there are many density perturbations along the line of sight, leading to most of the signal cancelling between over- and under-densities.
On small-scales there can be additional ISW contributions even in matter domination from non-linear growth of structure (the Rees-Sciama effect~\cite{ReesSciama}), however these are very small \cite{SeljakISW, SmithISW,Ferraro:2022twg}, so we focus on the linear contribution.

The correlation between the lensing potential and $\Delta T(\vnhat)_{\rm ISW}$ is very high ($\agt 0.9$), potentially making CMB lensing an excellent probe of the ISW signal. \JC{A detection of the \planck~lensing-ISW bispectrum was given in \cite{Planck:2015zfm}, and using temperature lensing cross-correlation in \cite{Ade:2015zua}}. The ISW can also be seen in cross-correlation with other large-scale structure probes, as first detected by Ref.~\cite{Boughn:2003yz} (see Ref.~\cite{Planck:2015fcm} for a review of subsequent results). CMB lensing has the nice property that for a given cosmology the amplitude  and redshift kernel are accurately predicted (no bias or source redshift uncertainty), and the signal can be reconstructed over most of the sky. Since the correlation is so high, CMB lensing also has most of the signal. For the foreseeable future, \planck\ observations are the only ones that can reconstruct lensing over the full sky~\cite[hereafter PL2018]{PL2018}, so the \planck\ lensing map will remain the best lensing probe of the large-scale ISW cross-correlation for some time. It is therefore worth trying to get the best reconstruction, and constructing a likelihood that can be used in cosmological parameter analysis of extended models. There has been no previously-published \planck\ ISW likelihood, so this is a new (if admittedly not very powerful) \planck\ product.

Previous \planck\ ISW cross-correlations results are extensively discussed in Ref.~\cite{Planck:2015fcm}.~\planck\ lensing-temperature cross-correlation spectrum results were given in \cite{Ade:2015zua}, and recently updated in the PR4 lensing analysis~\cite[hereafter PL4]{Carron:2022eyg}. The PR4 lensing analysis uses more optimal filtering to improve the lensing signal recovery, and also uses the new NPIPE (PR4) \planck\ CMB maps~\cite{Akrami:2020bpw} (which include more data from satellite repointing periods and improve many parts of the data processing). In this paper we use lensing maps from PL4 to construct an ISW likelihood, which we then use to constrain the monopole CMB temperature independently of the  COBE/FIRAS results~\cite{Fixsen:1996nj}.

\section{Modelling}
\begin{figure}
   \centering
   \includegraphics[width=\hsize]{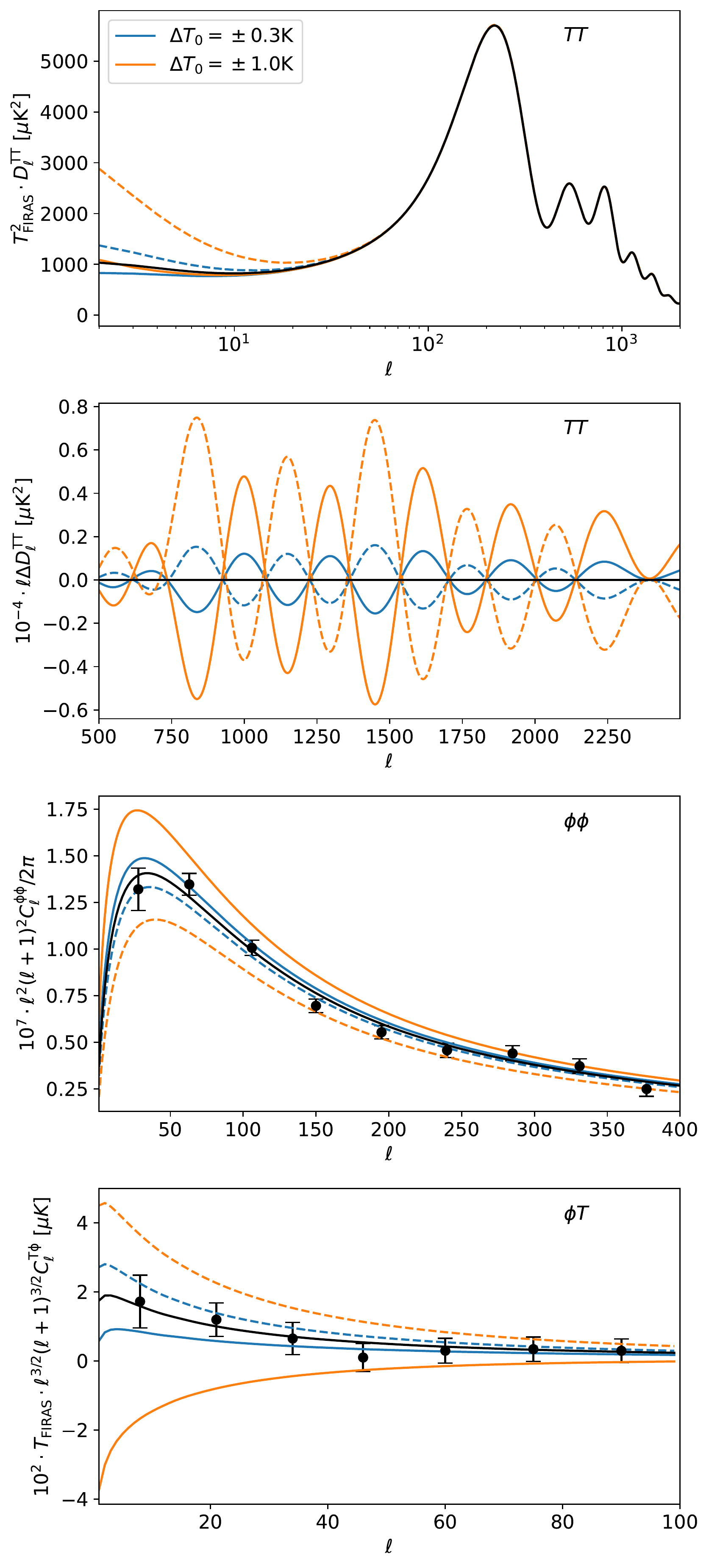}
      \caption{Change of the CMB temperature and lensing spectra along the main degeneracy line defined by constant $\omega_b/T_0^3, \omega_c/T_0^3, A_sT_0^{n_s - 1}$ and $\theta_\star$ and $n_s$, for varying background CMB temperature $T_0$ around $\Tfiras$. \JC{Solid (dashed) lines show positive (negative) $T_0$ increments}. The impact on the ISW-lensing $C_\ell^{\phi T}$ is much larger in relative terms than on the lensing spectrum. For \planck~noise levels, this results in the lensing and ISW-lensing spectra having almost equivalent constraining power on $T_0$ when considered independently, despite the much more precise measurement of the former. The blue and orange solid lines are obtained using CMB temperature values close to the PR3 and PR4 CMB $TT$ spectra best-fits.  The two lower panels also show the lensing and lensing-ISW PR4 data points used in this work. On the first two panels $D^{TT}_\ell$ is $\ell (\ell + 1) C^{TT}_\ell /2\pi$. The relative constraining power of these effects on the spectra can be seen in Fig.~\ref{fig:H0stoypdfs}.}
         \label{fig:dcldT0}
  \end{figure}

 \begin{figure}
   \centering
   \includegraphics[width=\hsize]{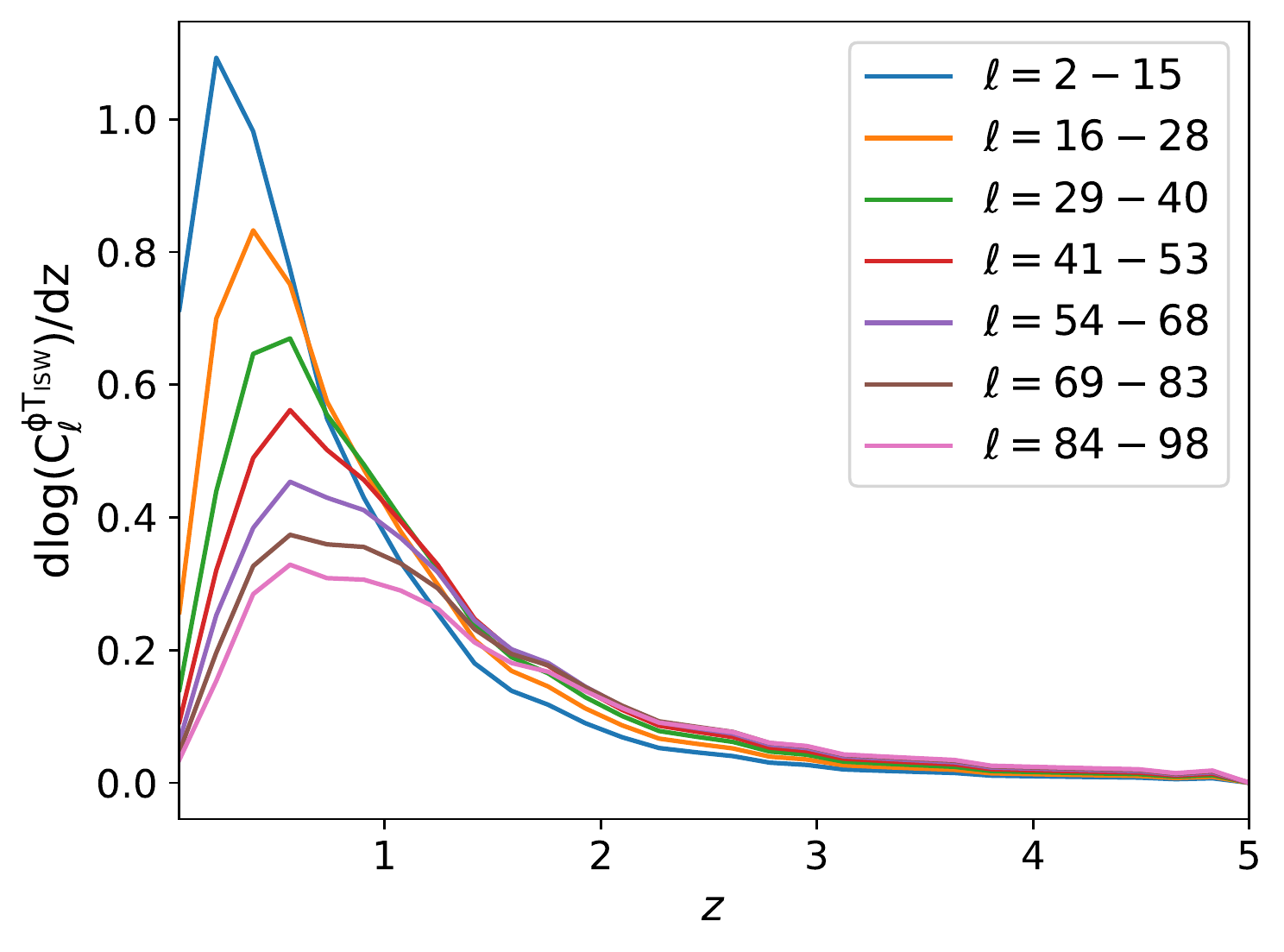}
      \caption{Per redshift contribution to $C_\ell^{\phi T_{\rm ISW}}$ within $\Lambda$CDM, for the amplitudes bins shown in the legend.
      }
         \label{fig:dCdz}
  \end{figure}
We aim to construct a likelihood based on the cross-correlation spectrum estimator $\hat C_\ell^{\phi T}$ between lensing reconstruction and the CMB temperature.
For \planck~noise levels, the covariance between the CMB lensing reconstruction and the CMB spectra is known to be very weak and can be neglected~\cite{Schmittfull:2013uea,Peloton:2016kbw}.
The main correlations to consider are therefore between $\hat C_\ell^{\phi T}$ and $\hat C_\ell^{T T}$, and $\hat C_\ell^{\phi T}$  and $\hat C_\ell^{\phi \phi}$. \JC{The CMB polarization E-mode, being correlated to the primordial temperature, could in principle be used to increase slightly the signal to noise of $\hat C_\ell^{\phi T}$~\cite{2009MNRAS.395.1837F}. The prospects are however modest (we forecast an improvement of at best $9\%$), and would require thorough understanding of the low polarization multipoles, where foregrounds and systematics are certainly more worrisome than in temperature. For these reasons we do not consider this possibility in this work.} For simplicity of use, we create difference likelihoods containing the additional ISW information, so that the new likelihoods can simply be combined with the standard full-resolution \planck\ likelihoods.
%as well as a joint $\hat C_\ell^{\phi T}$-$\hat C_\ell^{\phi \phi}$ likelihood including the lensing auto spectrum.
%We construct a $\hat C_\ell^{\phi T}$ as well as a joint $\hat C_\ell^{\phi T}$-$\hat C_\ell^{\phi \phi}$ likelihood.
%In order to be able combine with the official \planck~likelihoods in a consistent manner, we must take into account the correlation to the CMB spectra. This leaves us to consider in more details the covariances of the $TT$ spectrum to $\hat C_\ell^{\phi T}$ and $\hat C_\ell^{\phi \phi}$.
We first construct a joint $\{\hat C_\ell^{TT}$, $\hat C_\ell^{\phi T}$, $\hat C_\ell^{\phi \phi}\}$ likelihood at low multipoles ($2 \leq \ell \leq 100$). \JC{We model the contribution to the likelihood as a Gaussian $\propto e^{-\frac 12 \chi^2}$ with fixed covariance and discard the constant determinant normalization~\cite{Carron:2012pw}}. We write then
\begin{equation}\label{eq:chi2ISW}
	\chi^2 \equiv \chi^2\left(\hat C_\ell^{TT}, \hat C_\ell^{ \phi T}\right) - \chi^2\left(\hat C_\ell^{TT}\right) \textrm{ (for lensing-ISW only)}
\end{equation}
for combination  with the \planck~TT likelihood. For combination with both the \planck~TT and lensing likelihoods we instead have
\begin{equation}\label{eq:chi2ISWpp}\begin{split}
	&\chi^2 \equiv \chi^2\left(\hat C_\ell^{TT}, \hat C_\ell^{\phi T}, \hat C_\ell^{\phi\phi}\right) - \chi^2\left(\hat C_\ell^{TT}\right) - \chi^2\left(\hat C_\ell^{\phi\phi}\right).
\end{split}
\end{equation}
\JC{One can motivate these equations as follows: within our joint-likelihood model, $-\frac 12 \chi^2$ of Eqs.~\eqref{eq:chi2ISW} and \eqref{eq:chi2ISWpp} are the conditional probabilities $\ln p(\hat C_\ell^{\phi T} | \hat C_\ell^{TT})$ and $\ln p(\hat C_\ell^{\phi T} | \hat C_\ell^{TT},\hat C^{\phi\phi}_\ell)$ respectively. According to Bayes' theorem, their combination with the official \planck~likelihoods then gives the joint result including the new cross-correlation measurement. Although we assume Gaussianity for constructing the ISW-difference likelihood, the combination with full \planck\ low-$\ell$ likelihood accounts more accurately for the non-Gaussianity of the CMB $TT$ spectrum at low multipoles.}
To build our likelihoods we use the lensing reconstruction maps, as well as Wiener-filtered CMB maps that are obtained by the lensing   reconstruction pipeline as input to the lensing map estimators.
If the instrument noise and CMB are close to Gaussian with accurately-known spectra, the spectrum of the Wiener-filtered map is a sufficient statistic for the CMB likelihood. The noise and foreground model are not accurate in practice, however the temperature noise is very small on the largest scales, and foregrounds can be cleaned, so we ignore these differences. Using a fixed fiducial $C_\ell^{TT, \rm fid}$ spectrum for the filtering may be slightly sub-optimal, but this resulting `quadratic maximum likelihood' (QML) estimator~\cite{Tegmark:1996qt} can still be used to construct an unbiased Gaussian likelihood.

In Sec.~\ref{sec:datavector} we first discuss the construction of the $\hat C_\ell^{TT}$ and $\hat C_\ell^{\phi T}$  data vectors and the modelling of their predictions. Their variances and covariances (also to $\hat C_\ell^{\phi\phi}$) are discussed in Sec.~\ref{sec:covariances}. Plots of the relevant covariance matrices are relegated to the end of the paper. We do not discuss the $\hat C_\ell^{\phi\phi}$ data vector, which is exactly the same as in PL4; it is built using the most precise, inhomogeneously-filtered, $\kappa$-filtered \cite{Mirmelstein:2019sxi} lensing maps. For simplicity of the modelling, when building $\hat C_\ell^{\phi T}$ we instead use the PL4 lensing maps built with the 2018 \planck~lensing pipeline, which uses homogeneous noise filtering at a slight cost in signal to noise. \JC{We use the minimum variance (MV) quadratic estimator (QE) reconstructions, that combine the temperature and polarization QEs in a way that is approximately optimal.} Our new $\hat C_\ell^{\phi T}$ data points can be seen on the lowest panel of Fig.~\ref{fig:dcldT0}, and formally give a 4$\sigma$ detection of a non-zero signal consistent with our fiducial \planck~FFP10\footnote{\url{https://github.com/carronj/plancklens/blob/master/plancklens/data/cls/FFP10_wdipole_params.ini}}  cosmology (shown as the black solid line). Fig.~\ref{fig:dCdz} shows how the signal in each bin depends on redshift.
\subsection{Data vectors}\label{sec:datavector}
  \begin{figure*}
   \centering
   \includegraphics[width=\hsize]{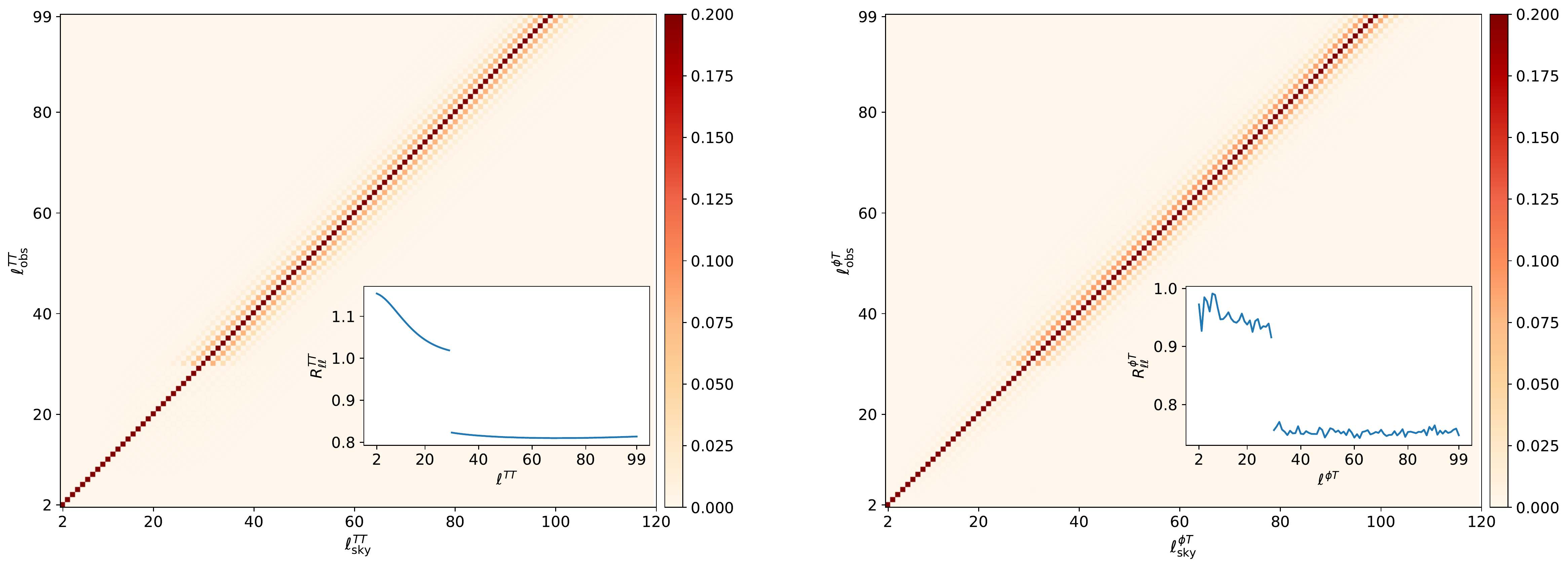}
      \caption{\emph{Left panel}: Unbinned Wiener-filtered temperature spectrum fiducial amplitude coupling matrix (see Eq.~\ref{eq:ttcoupling}), for the multipole ranges  $2\le \ell_{\rm obs} \le 99$ and $2\le \ellsky \le 120 $ (by construction our filtering has vanishing response to the CMB dipole), obtained as described in the main text. For $\ell < 30$, results are built on a larger sky fraction ($86\%$), resulting in almost perfectly diagonal responses and covariances. For plotting the matrix is rescaled by its diagonal elements (of order unity, shown in the inset), as for a cross-correlation matrix. \emph{Right panel}: Same for the $\phi T$ spectrum amplitude estimates, defined in Eq.~\eqref{eq:RPT}, where some residual Monte-Carlo noise remains visible.\label{fig:ttcoupling}}
  \end{figure*}
\begin{figure}
   \centering
   \includegraphics[width=\hsize]{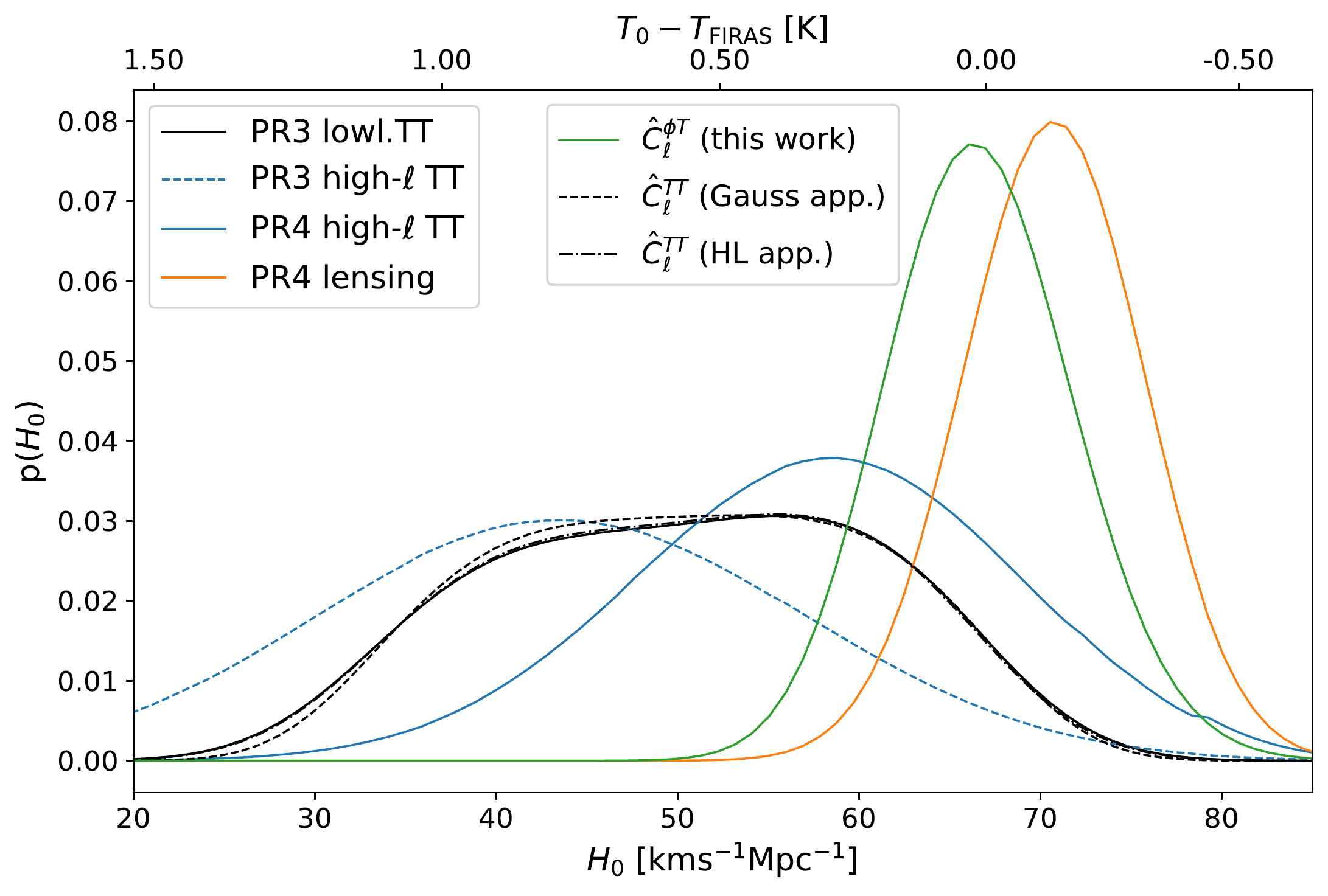}

      \caption{Illustration of the relative constraining power of different parts of \planck~data on the $H_0$-$T_0$ degeneracy (see also Fig.~\ref{fig:dcldT0}). The curves are obtained by evaluating likelihoods in the toy one-parameter model defined by constants $\omega_b/T_0^3, \omega_c/T_0^3, A_sT_0^{n_s - 1}$ and $\theta_\star$ (as well as fixed $n_s$ and $\tau$), which captures well the qualitative behavior of the full \LCDM+$T_0$ results (see Fig.~\ref{fig:T0_mcmc} and Fig.~\ref{fig:PR4mcmc}). The black curve (\planck~low-$\ell$ TT) captures the large-scale ISW effect.  The high-$\ell$ CMB constraints (comparable to that of the low-$\ell$) come from the differential lensing smoothing effect, and differ somewhat significantly between PR4 (blue) and PR3 (dashed blue), owing to the larger sky area used for the PR4 \camspec~likelihood compared to the PR3 \plik~likelihood. The orange curve is obtained with the PR4 lensing power spectrum alone. The green curve shows the constraint from the lensing-ISW data alone, and is new to this work. The black dashed and dot-dashed \JC{(`HL',  Ref.~\cite{Hamimeche:2008ai})} lines are approximations to the \planck~low-$\ell$ TT likelihood that we build and use to take into account the covariance of the $\phi T$ and $TT$ spectra as discussed in the main text. The centre of the approximately flat region of these posteriors corresponds to $\Omega_\Lambda$ changing sign and becoming increasingly negative at high CMB temperature. In the full $\Lambda$CDM + $T_0$ parameter space, additional degeneracies slightly reduce the statistical power of the lensing spectrum.
               \label{fig:H0stoypdfs}}
  \end{figure}

The public \planck~temperature-based likelihoods at $2\leq \ell \le 29$ are built differently from those at higher multipoles, and on different sky areas. A large fraction of the signal to noise on $\hat C_\ell^{\phi T}$ comes from this low-$\ell$ range, but some part of signal extends at higher multipoles (about 50\% of the SN is located below $\ell = 10$, and $5\%$ above $\ell = 75$). In order to model more accurately the covariances with $\hat C_\ell^{TT}$, we use two temperature maps to build our bandpowers. For $2\leq \ell \leq 29$, we use temperature maps built on the same mask as the low-$\ell$~\planck~TT likelihood, with $f_{\rm sky} \sim 86\%$. Above $\ell = 30$, we construct all bandpowers on the \planck~PR4 lensing mask, which covers $67\%$ of the sky (the lensing masks differ to a very minimal extent between the PL3 or PL4 analyses). In this latter case, we neglect the slight differences in sky area and methodology used for the high-$\ell$ TT likelihoods (described below). The differences are expected to be small, with the TT signal to noise on the range $30\le \ell \le 100$ matching to percent level that of the PR4 high-$\ell$ TT that we use ($67.8$ compared to $68.9$). In any case, our $TT$-bandpowers only serve to model the small covariance to $\hat C_\ell^{\phi T}$, which is at most  0.1 for $\ell \ge 30$. The approximations we make on the higher multipole range are therefore not critical (in fact, none of this $\ell$-range has any impact on the results on the internal constraint on the CMB temperature  shown in this paper). On the low multipole range, our construction of the $TT$-likelihood matches the public likelihood very well (at least for our usage later on, as can be seen from the black lines in Fig.~\ref{fig:H0stoypdfs}). On the entire multipole range, the lensing maps are built on the lensing mask.
\\ \indent
The first step of our analysis pipeline is to build Wiener-filtered CMB maps ($\TWF_{\ell m}, E^{\rm WF}_{\ell m}$, $B^{\rm WF}_{\ell m}$). These maps are used for the construction of the lensing map and spectrum, and the filtered temperature is also directly used for the ISW-lensing cross-correlation with the large-scale lensing map, and to build the covariance to the TT auto-spectrum. On PR3 data we use the official foreground-cleaned SMICA maps, and for PR4 data the same SMICA maps that were built for PL4, to which we refer for details on their construction. The same Wiener-filtering procedure is applied to PL3 and PL4, using conjugate gradient descent. In the case of temperature-only, and using the notation of those papers, the equation to be solved is
\begin{equation}\label{eq:TWF}
	\TWF = C^{TT, \rm fid}\mathcal T^\dagger \Cov^{-1} T^{\rm dat}.
\end{equation}
The fiducial covariance model $\Cov$ always uses a fiducial transfer function model $\mathcal T$ built out of an isotropic beam of $5'$ together with the pixel window function, and an homogeneous noise level of $32 \mu \rm{K}$-amin across the unmasked area, with the exception of the maps used to construct $\hat C_\ell^{\phi\phi}$ which are built as described in PL4 and account for noise inhomogeneity.

 \subsubsection{$TT$-data}\label{sec::ttdata}
From the $T^{\rm WF}_{\ell m}$ filtered maps, we first build fiducial amplitude estimates. Using the available FFP10 noise-only simulations of our foreground-cleaned maps (for PR4, these also include large-scale foregrounds residuals), we estimate a noise contribution $\hat N_\ell$ to the auto-spectrum of the filtered data map by filtering them and averaging their spectra, and then build
  \begin{equation}\label{eq:ttamplitude}
 	\hat A_\ell^{TT} C_\ell^{TT, \rm fid}\equiv  \left( \frac{1}{ f_\ell^{TT}(2\ell + 1)}\sum_{m=-\ell}^{\ell} \left|T^{\rm WF}_{\ell m}\right|^2 \right) - \frac{\hat N^{\rm WF}_{\ell}}{f_\ell^{TT}} .
 \end{equation}
The factor $f^{\rm TT}_L$ applies a preliminary crude isotropic normalization, accounting for masking and the Wiener filter,
 \begin{equation}
 	f_\ell^{TT} \equiv \fsky \left(\frac{C_\ell^{TT, \rm fid}}{C^{TT, \rm fid}_\ell + N^{TT, \rm fid}_\ell} \right)^2.
 \end{equation}
In this equation $ N^{TT, \rm fid}_\ell$ is the white noise prediction of our fiducial covariance model. Both $\hat N^{\rm WF}$ and $N^{TT, \rm fid}$ are tiny corrections and largely irrelevant on all scales considered for the cross-correlation to the lensing.
The amplitude estimator
$\hat A^{\rm TT}_\ell$ is close to unbiased, matching expectation across simulations to about 5\% on most scales and up to 15\% on the very smallest multipoles. This mismatch is caused by the residual mode-coupling still present after Wiener filtering. We define the response matrix $\Rtt_{\ell \ellsky}$ to the true CMB spectrum $C_{\ellsky}^{ TT}$ as
 \begin{equation}\label{eq:ttcoupling}
 \av{\hat A^{TT}_\ell} = \sum_{\ellsky} \Rtt_{\ell \ellsky} \left(\frac{C_{\ellsky}^{TT}}{C_{\ellsky}^{TT, \rm fid}} \right).
 \end{equation}
We get the response matrix as follows. Let $F_{\ell m}^{\ellsky \msky}$ be the matrix representation of the linear Wiener-filtering operation, connecting the Wiener-filtered CMB $T_{\ell m}$ mode to the sky mode $T_{\ellsky \msky}$. In terms of the fiducial covariance matrix model of Eq.~\eqref{eq:TWF}, $F$ may be written
\begin{equation}\label{eq:Fmat}
	F \equiv C^{TT, \rm fid}\mathcal T^\dagger \Cov^{-1} \mathcal T.
\end{equation}
From its definition,  Eq.~\eqref{eq:ttcoupling}, together with Eq.~\eqref{eq:ttamplitude}, the response matrix is directly proportional to %its  isotropized square,
 \begin{equation}
 	\Rtt_{\ell \ellsky} \propto\sum_{m,\msky} \left|F_{\ell m}^{\ellsky, \msky}\right|^2.
 \end{equation}
 For all purposes in this paper, $\ell$ is at most $100$, and the coupling extends only across a small range of multipoles. For these reasons the matrix $F$ remains small enough that it can be explicitly calculated via brute force calculation: Wiener-filtering an input map with a single non-zero $(\ellsky, \msky)$ mode directly gives the corresponding entire matrix row. There are $(\ell_{\rm max} + 1)^2$ modes up to multipole $\ell_{\rm max}$. The entire matrix can thus be obtained by filtering $(\ell_{\rm max} + \Delta \ell + 1)^2$ maps, where $\Delta\ell$ is a buffer accounting for the couplings to modes smaller than $\ell_{\rm max}$.  Since all modes are degree-scale or larger, for this we can use a degraded version of the filter working at a coarser pixel resolution than the 1.7' of the native \planck~maps. We used four times larger pixels and a very generous $\Delta \ell = 100$.
 The unbinned coupling matrix is shown on Fig~\ref{fig:ttcoupling}.
 Due to the approximate symmetry of the mask with respect to the galactic equator, the non-diagonal elements of $\Rtt$ are most prominent for $|\ell- \ellsky| = 2$, but always very small. On the \planck~lensing mask, we see almost constant couplings of size  $9\%, 3\%$ and $0.2\%$ relative to the diagonal for $|\ell- \ellsky| = 2 , 4$ and $1$ respectively. On the larger sky area used below $\ell < 30$, the matrix is almost perfectly diagonal.

  We use $\Rtt$ not to undo the couplings in our amplitude estimates (which would require inverting the matrix), but rather to correct the prediction of the amplitude; this choice does not affect the information content of the spectrum likelihood. On the FFP10 simulation suite, the estimates are then biased at most by a tenth of an error bar.
 \subsubsection{$\phi T$-data}\label{sec:pTresp}

 Similarly, from the filtered temperature multipoles and lensing reconstruction estimator $ \hat \phi_{\ell m}$ we first build fiducial amplitudes
  \begin{equation}
 	\hat A_\ell^{\phi T}  C_\ell^{\phi T, \rm fid} \equiv \frac{1}{f^{\phi T}_{\ell}(2\ell + 1)}\sum_{m=-\ell}^{\ell} \hat \phi_{\ell m}T^{\rm WF,\dagger}_{\ell m},
 \end{equation}
 with
 \begin{equation}\label{eq:flpt}
 	f_\ell^{\phi T} \equiv \fsky \left(\frac{C_\ell^{ TT, \rm fid}}{C^{TT,\rm fid}_\ell + N^{TT, \rm fid}_\ell} \right).
 \end{equation}
 Compared to the $TT$ case, it might appear less natural to use an amplitude defined with respect to a fiducial $C_\ell^{\phi T, \rm fid}$ here, which can be zero in some models. However, the fiducial spectrum, as well as any prefactor like $f_\ell^{\phi T}$, cancels out in the Gaussian likelihood and hence does not affect final results.
Since we only consider large scales, the temperature field entering the cross-correlation has negligible lensing contribution and can be treated as unlensed. The correlation with the lensing quadratic estimator is therefore completely dominated by contractions proportional to $C_\ell^{\phi T}$ and the lensing response functions are chosen to make the cross-correlation estimator non-perturbatively unbiased on the full sky~\cite{Lewis:2011fk}. We may therefore write the general estimator response as

\begin{equation}\label{eq:RPT}
	\av{\hat A^{\phi T}_\ell} = \sum_{\ellsky}\Rpt_{\ell \ellsky} \left( \frac{ C^{\phi T}_{\ellsky}} {C_{\ellsky}^{\phi T, \rm fid}} \right),
\end{equation}
for some matrix $\Rpt$. In contrast to $\Rtt$, $\Rpt$ now has a dependence on the cosmological model (though only weakly so) through the lensing QE estimator response. We account for this in our likelihood in a way described further below. We first obtain an unbinned matrix $\Rpt$ in the fiducial cosmological model in the following manner.\\
We produce CMB-only simulations in pairs, where the members of each pair share the same unlensed $T$ and $E$ maps. The lensing potentials deflecting these unlensed CMB are also very similar, with the difference that the first pair member has the expected (small) cross-correlations $C_\ell^{\phi T}$ and $C_\ell^{\phi E}$, while for the second they have been set to zero. We then perform the \planck~MV QE reconstruction on both maps, resulting in $\hat \phi^{w. \rm ISW}$ and $\hat \phi^{n. \rm ISW}$ respectively, and obtain an estimate of the response matrix through the cross-spectra
\begin{equation}
	\hat\Rpt_{\ell \ellsky}\propto \sum_{m, \msky}\left(\hat \phi_{\ell m}^{w. \rm ISW} - \hat \phi_{\ell m}^{n. \rm ISW} \right) F^{\ellsky \msky}_{\ell m} T^{\rm unl}_{\ellsky \msky},
\end{equation}
where $T^{\rm unl}$ is the unlensed temperature of the pair, and $F$ the dense filtering matrix calculated in the previous subsection (see Eq.~\eqref{eq:TWF}). Using this QE difference greatly reduces the Monte-Carlo noise of this estimate, by cancelling to a very high degree the lensing reconstruction noise as well as the mean-field of the signal-carrying  $\hat \phi^{w. \rm ISW}$, and provides good estimates of all of the matrix entries. As for $\hat A^{TT}$, we use this matrix to forward-model the couplings in our amplitude predictions.

  It is well known that in addition to the main dependency on the lensing spectrum, the lensing QE gets an additional model dependence through its normalization: on an isotropic sky, we may write the QE signal part to good accuracy as
 \begin{equation}\label{eq:respratio}
 	\hat \phi_{\ell m} \propto \frac{\mathcal R_{\ell}(\theta)}{\mathcal R_\ell(\theta^{\rm fid})} \phi_{\ell m},
 \end{equation}
 where $\mathcal R(\theta^{\rm fid})$ is the (arbitrary) normalization that was applied to the estimate, and $\mathcal R(\theta)$ the true sky lensing response. This is almost always a very small effect, since the CMB spectra are known empirically to a very high accuracy already, leaving little wiggle room for significant variations in the response in most models. The dependency enters exclusively through the CMB spectra, and is linear in them. We include it in our likelihood by precomputing the matrices $d \ln \mathcal R_{\ell} / d\ln C_{\ell'}^{XY}$ for $XY \in (TT, TE, EE)$, allowing us to recalculate quickly the isotropic response for each point in a Monte Carlo Markov chain (MCMC) parameter space. We then rescale the prediction by the response ratio of Eq.~\ref{eq:respratio}. In doing so we neglect the mask-induced couplings for the purpose of the parameter dependence, which is perfectly adequate since the couplings are themselves a few percent level correction already.

\subsection{Covariances}\label{sec:covariances}
In this section we describe how we build the various covariance matrix blocks. We build these blocks in the same way, but using temperature maps built on two different masks for $2 \leq \ell \leq 29$ and $30 \leq \ell \leq 100$ respectively. Selected figures with unbinned covariances are given in the appendix. As discussed at the beginning of this section, the covariance between $\hat C_\ell^{\phi\phi}$ and $\hat C_\ell^{TT}$ can be neglected and is not discussed here.
\subsubsection{$TT$ - $TT$  covariance}
In addition to the empirical covariance of the spectra from simulations, we also built a couple of improved estimates to the covariance, showing that, for all practical purpose, the non-idealities of the CMB maps (apart of masking) and the noise contribution can be safely neglected. The dense filtering matrix $F$ of section~\ref{sec::ttdata}, in conjunction with the input CMBs, allows us to test for the importance of non-idealities in the CMB and noise FFP10 simulations. To do this, we improve the convergence rate of the empirical covariance by subtracting a covariance estimate built from the input CMBs and the dense filtering matrix, and adding the exact analytic mean of this estimate. This subtracts most of the ideal-CMB realization-dependent variance, giving off-diagonal coefficients that are smaller by about a factor of 10 or so. The resulting covariance matrix accelerated in this way shows no significant feature at all, except for the expected mode-coupling. Our prediction of the covariance from the dense matrix seems to be an excellent fit to the empirical matrix, and is used for our covariance in what follows.

\subsubsection{$\phi T$ - $TT$ covariance}
Since the large-scale temperature modes are effectively unlensed, and that the lensing map is built from high multipoles only, the expected covariance only comes from the mode-coupled disconnected Gaussian signal proportional to the product of $\phi T$ and $TT$ sky spectra. The cross-correlation of lensing to temperature sharply decays with multipole as the ISW signal decays, so this covariance should only be relevant on the very largest scales. To get a more precise unbinned estimate of this Gaussian covariance than just the naive empirical covariance, we may \JC{proceed as follows: according to Wick's theorem, the Gaussian part consists of the product of the two pairs (neglecting scaling factors and constants for simplicity)
\begin{equation}\label{eq:covPTTT}\begin{split}
	\textrm{Cov}[\hat A^{\phi T} \hat A^{TT}]_{\ell_1 \ell_2} &\propto \sum_{m_1, m_2} \av{\hat \phi_{\ell_1 m_1}\TWFd_{\ell_2 m_2}}\av{\TWF_{\ell_1 m_1}\TWFd_{\ell_2 m_2}} %\\ &= \sum_{m_1, m_2}\av{\hat \phi_{\ell_1 m_1} K_{\ell_1m_1}^{\ell_2 m_2}\TWFd_{\ell_2 m_2}}.
	\end{split}
\end{equation}
The pairing of large-scale temperatures on the right-hand side contains so little noise that we can use an analytic formula for it, assuming the noise model in the filter matches that of the data. Under this assumption, and defining the matrix $K$ as
\begin{equation}
	K_{\ell 1 m_1}^{\ell_2 m_2} \equiv \av{\TWF_{\ell_1 m_1}\TWFd_{\ell_2 m_2}},
\end{equation}
we then have
\begin{equation}
		K = C^{TT, \rm fid}\mathcal T^\dagger \Cov^{-1} \mathcal T C^{TT, \rm fid} = F C^{TT, \rm fid}.
\end{equation}
The matrix $K$ can computed by brute force from $F$, defined in Eq.~\eqref{eq:Fmat}. Hence we may write
\begin{equation}\label{eq:covPTTT}\begin{split}
	\textrm{Cov}[\hat A^{\phi T} \hat A^{TT}]_{\ell_1 \ell_2} &\propto \sum_{m_1, m_2}\av{\hat \phi_{\ell_1 m_1} K_{\ell_1m_1}^{\ell_2 m_2}\TWFd_{\ell_2 m_2}}.
	\end{split}
\end{equation}
}
For a single Monte-Carlo simulation, and each $\ell_1, \ell_2$, Eq.~\ref{eq:covPTTT} is now in the form of matrix-vector multiplications, which can easily be performed. Such an estimate will contain little Monte-Carlo noise. Fig.~\ref{fig:cov_TTPT} shows the empirical covariance estimate using the 480 FFP10 simulations, and our Gaussian covariance estimate, which seems be a perfectly adequate model.

\subsubsection{$\phi T$-$\phi T$ covariance}
Here, we also we assume the disconnected contractions provide a good model. There are two such terms, one proportional the product of the $\hat T \hat T$ and $\hat \phi \hat \phi$ auto-spectra, and the other the square of $\hat \phi \hat T$,
\begin{equation}\begin{split}\label{eq:covPTPT}
	\textrm{Cov}[\hat A^{\phi T} \hat A^{\phi T}]_{\ell_1 \ell_2} &\propto \sum_{m_1 m_2} \av{ \hat\phi_{\ell_1 m_1} \hat \phi^\dagger_{\ell_2 m_2} }\av{ \TWFd_{\ell_1 m_1} \TWF_{\ell_2 m_2} } \\&+\sum_{m_1 m_2} \av{ \hat\phi_{\ell_1 m_1} \TWFd_{\ell_2 m_2} }\av{\TWF_{\ell_1 m_1}  \hat \phi^\dagger_{\ell_2 m_2}  }.
	\end{split}
\end{equation}
The first term strongly dominates almost everywhere. To isolate the contributions, we proceed as follows. For the first term in Eq.~\eqref{eq:covPTPT} we use the form
\begin{equation}\label{eq:covPTPT_PPTT}
	\textrm{Cov}[\hat A^{\phi T} \hat A^{\phi T}]_{\ell_1 \ell_2} \ni \av{ \hat\phi_{\ell_1 m_1}  K_{\ell_1 m_1}^{\ell_2 m_2}\hat \phi^\dagger_{\ell_2 m_2} },
\end{equation}
similar to Eq.~\eqref{eq:covPTTT}, where we average over lensing estimates from the FFP10 simulations. To obtain the second contribution, we use the ISW-paired noise-free CMB simulations of Sec.~\ref{sec:pTresp} to build
\begin{equation} \begin{split}
\Cov \left[ d\hat A^{\phi T}_{\ell_1}  d\hat A^{\phi T}_{\ell_2} \right] - \av{ \delta  \hat\phi_{\ell_1 m_1}  K_{\ell_1 m_1}^{\ell_2 m_2}\:\hat \delta \hat\phi^{\dagger}_{\ell_2 m_2}  },
\end{split}
\end{equation}
with $\delta \hat \phi \equiv  \hat \phi_{\ell m}^{w.\rm ISW} -\hat \phi_{\ell m}^{n. \rm ISW} $ and (we are suppressing throughout $(2\ell + 1)$ and other prefactors to avoid cluttering)
\begin{equation}
	d\hat A^{\phi T}_\ell =\sum_m \TWF_{\ell m}\delta \hat \phi_{\ell m}^\dagger.
\end{equation}
This term is at most a percent-level correction to that in Eq.~\eqref{eq:covPTPT_PPTT} on the lowest multipoles, and could have been safely ignored. The error bars for $\hat A_\ell^{\phi T}$ calculated in this way accurately match the empirical errors from the FFP10 simulation suite, as shown in Fig.~\ref{fig:covs} (right panel).
 %----
 \subsubsection{$\phi T$-$\phi \phi$ covariance}
 Due to the separation of scales between the modes used for lensing and for $\TWF$, the covariance of $\hat A^{\phi T}$ with $\hat A^{\phi \phi}$ is expected to come from the ISW signal itself. The Gaussian isotropic approximation predicts a positive cross-correlation of size $\sim 0.3$ at the quadrupole down to a percent level for $\ell \sim 100$. To obtain a good unbinned covariance model from the FFP10 simulations, we reduce primordial CMB variance by estimating the covariance using the covariance of $\hat A^{\phi T^{\rm ISW}}$, where
 \begin{equation}
 	T^{\rm ISW}_{\ell m} = \frac{C_\ell^{\phi T}}{C_\ell^{\phi\phi}}\phi_{\ell m}
 \end{equation}
 is the input ISW-signal part of the simulated temperature map. We use our dense filtering matrix for this purpose. The resulting covariance matrix has much lower Monte-Carlo noise and appears almost diagonal, as can be seen on Fig.~\ref{fig:cov_PPTP}.
%---T0
\section{Constraints on $T_0$}
\begin{figure*}
   \centering
   \includegraphics[width=0.49\hsize]{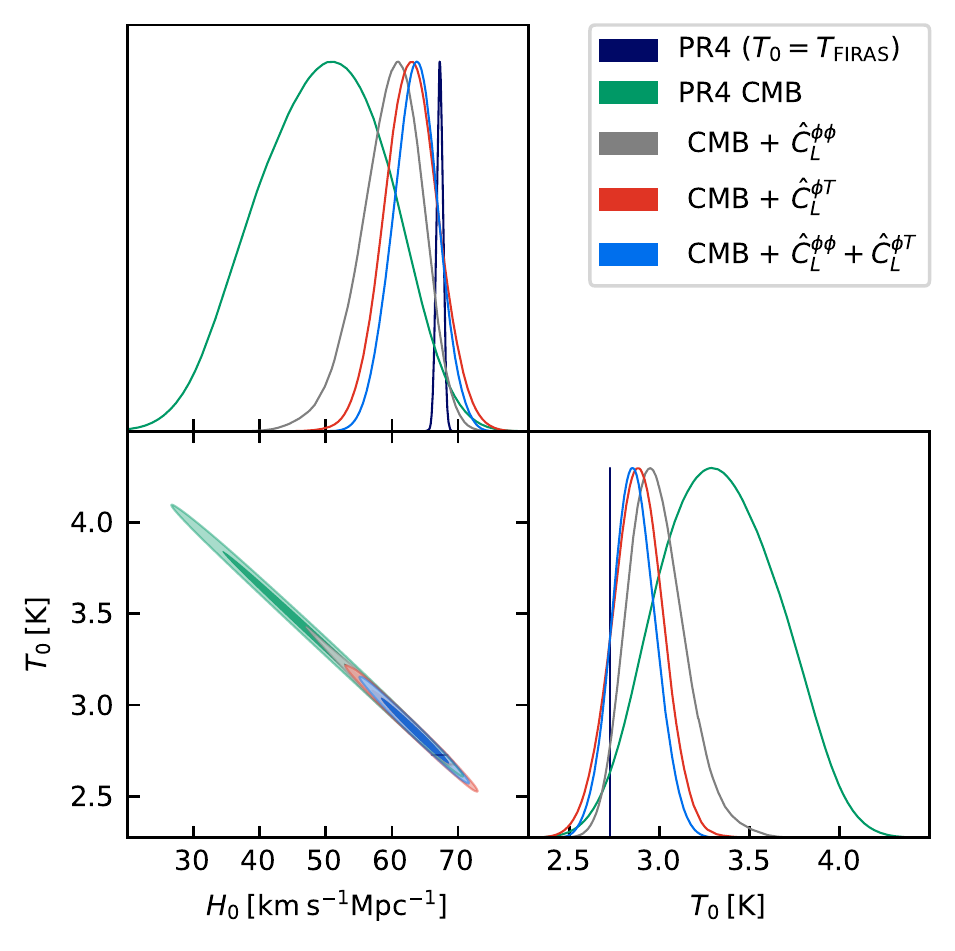}
   \includegraphics[width=0.49\hsize]{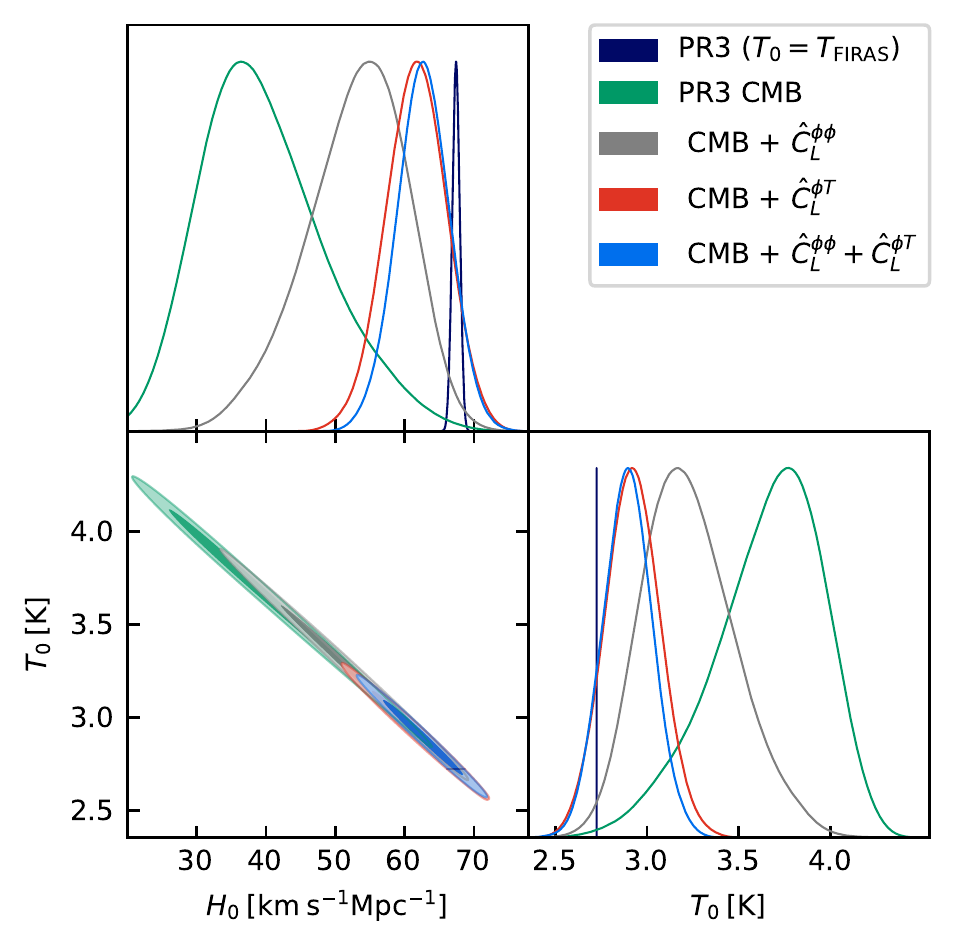}
   \caption{\label{fig:T0_mcmc}\emph{Left panel}: Posteriors on $H_0$ and $T_0$ using \planck~PR4 data in our \LCDM+$T_0$~MCMC chains, with and without the inclusion of the lensing and lensing-ISW data. Dark blue shows the fixed-temperature \LCDM~ results, including CMB and lensing data, for comparison. See Table~\ref{table:T0H0} for summary statistics and Fig.~\ref{fig:PR4mcmc} for the full parameter set constraints. \emph{Right panel}: Same for the PR3 data, with a somewhat different CMB-only constraint due to its greater preference for more lensing smoothing in the temperature spectrum.}
\end{figure*}
The background CMB temperature today, $T_0$, is usually fixed in cosmological analyses because it has been measured with tight error bars by the FIRAS instrument, achieving $T_0=(2.7255\pm 0.0006){\rm K}$~\cite{Fixsen:2009ug} when combined with WMAP data. This measurement uncertainty is sufficiently small that for current data marginalizing over it affects parameter constraints at a negligible level.
The FIRAS measurement remains the only measurement of $T_0$ at this precision, though there are previous measurements of comparable precision~\cite{PhysRevLett.65.537}. Subsequent observations have calibrated using the FIRAS result, so \planck\ measures $\Delta T/T$ and then scales the results to be reported in units of $\Tfiras\equiv 2.7255 {\rm K} $ without giving any direct temperature measurement.
It is therefore interesting to consider what happens if we do not impose the $T_0$ constraint, and we now describe how our new ISW likelihood can be used to constrain the CMB temperature independently.

In a homogeneous and isotropic cosmology, the CMB temperature scales $T\propto 1/a$, where $a$ is the scale factor, and so will appear different to observers at different times. The CMB temperature $T_0$ can then be thought of as parameterizing \emph{when} we are in this cosmology. Clearly a range of temperatures are consistent with exactly the same underlying evolution, just with different measured values of the Hubble parameter and a different scale factor at the time of observation. Since recombination happens at a fixed known temperature, the comoving angular diameter distance to last scattering also changes, because for lower observed temperatures the CMB is more distant. This means that observers at different times will see identical CMB acoustic peak structures, but the angular scale will be shifted to smaller scales at later times.

Within the framework of \LCDM~cosmologies, models related by differing values of the cosmological constant have the same early-universe physics, but different distance-redshift relations, so a shift in angular scale can also be compensated by a change in the cosmological constant while keeping nearly identical early-universe physics.
From observations of the linear CMB, there is therefore a very near parameter degeneracy between $T_0$ and $\Omega_\Lambda$ (and hence $H_0$, the `geometric degeneracy'). This is illustrated in Fig.~\ref{fig:dcldT0}, and discussed further in Refs.~\cite{Ade:2015xua,Ivanov:2020mfr,Wen:2020txi}.
The acoustic peak structure of the linear CMB power spectrum therefore gives almost no information about the CMB temperature.

This geometric degeneracy is broken on the small-scale anisotropies by CMB lensing, though only weakly. It is also broken to a comparable degree by the effect of the ISW signal on the large-scale temperature, which changes significantly with the change in cosmological constant required to keep the angular acoustic scale fixed as the CMB temperature varies. The CMB temperature therefore shows up as a strong dependency of the lensing-ISW cross-correlation spectrum, as shown in Fig.~\ref{fig:dcldT0}. The different mapping between redshift and time also affects the reionization signal, though for \planck\ this effect cannot be separately distinguished without knowing the true redshift evolution of reionization.
Here we focus on the large-scale ISW signal, and see whether our new likelihood can constrain the CMB temperature without using any non-\planck\ data.

We assume a base $\Lambda$CDM cosmology and follow the notation, assumptions and priors of Ref.~\cite{PCP2018}. We use \CAMB\footnote{\url{https://camb.info}}~\cite{Lewis:1999bs} to compute the theoretical power spectra and \Cobaya\footnote{\url{https://github.com/CobayaSampler/cobaya/}}~\cite{Torrado:2020dgo} to sample cosmological parameters with MCMC. Both these codes self-consistently handle varying the true CMB background temperature while allowing data constraints to be fixed in units of the FIRAS CMB temperature $\Tfiras$.
In \CAMB\ we use the {\tt Recfast} recombination model~\cite{Seager:1999km,Wong:2007ym}, generalized in \CAMB\ 1.3.6 to scale consistently with CMB temperature.
We use the \planck\ PR4 NPIPE TTTEEE CMB likelihood of Ref.~\cite{Rosenberg:2022sdy}, together with the \planck\ 2018 (PR3) low-$\ell$ temperature and $EE$ polarization likelihoods~\cite{PCP2018}.
The $EE$ likelihood mainly constrains the optical depth, with little dependence on the exact shape of the reionization history producing the polarization signal, and hence in itself does not help to break the $T_0$ degeneracy. For comparison we also show results using the official PR3 \planck\ TTTEEE {\sc plik} likelihood (which uses less sky area than Ref.~\cite{Rosenberg:2022sdy}, as well as different foreground and other modelling).

We sample $\omega_b/\bar T_0^3, \omega_c/\bar T_0^3, A_s \bar T_0^{n_s - 1}$ (with $\bar T_0 \equiv T_0/T_{\rm FIRAS} $ and $\omega_x \equiv \Omega_x h^2$), together with $n_s$, $\theta_\star, T_0$, and obtain $H_0$ as derived parameter. In this parameter space we do not constrain $\Omega_{\Lambda}$ to be positive, in contrast to Ref.~\cite{Ivanov:2020mfr}. Our results are shown on Figs.~\ref{fig:PR4mcmc} and \ref{fig:PR3mcmc}, with the relevant $H_0$-$T_0$ subspace reproduced on Fig.~\ref{fig:T0_mcmc}. We obtain chains using the CMB spectra-only likelihood, and with lensing and lensing-ISW alone and in combination, as well as one reference PR4 and PR3 \LCDM~\planck~chain where $T_0$ is fixed to the FIRAS value. For the reasons explained above, none of the \LCDM\ early-universe physical parameter constraints change significantly compared to that reference case, but $H_0$ becomes largely unconstrained. The PR4 CMB-spectra prefer a lower $H_0$, and, as visible on Fig.~\ref{fig:T0_mcmc}, and the PR3 spectra an even lower value. This is due to the well-known shape of the residuals of the high-$\ell$ $TT$ \planck~spectrum, preferring a higher level of lensing-like peak smoothing in \LCDM~\cite[Fig.~24]{PCP2018}, which is also achievable with a higher CMB temperature in \LCDM + $T_0$~(second panel of Fig.~\ref{fig:dcldT0}). High CMB temperatures can remain acceptable to the low-$\ell$ TT data, and (to a smaller extent) to the lensing auto-spectrum data, but eventually give a very strongly negative ISW-lensing signal (negative cosmological constant) that is ruled out by our $\hat C_\ell^{\phi T}$ measurements. For PR3 and PR4, the lensing-ISW data constraining power outperforms that of the lensing spectrum.

Simple summary statistics are listed in Table~\ref{table:T0H0}.
Combining all of our spectra, from PR4 data we find the $68\%$ confidence limits
\begin{equation}
\begin{split}
T_0 &= (2.86\pm 0.12)\:\rm{K} \\ %\quad (\planck\:PR4) \\
H_0 &= (63.5\pm 3.4) \:\rm{km \:s^{-1} Mpc^{-1}}.
\end{split}
\end{equation}
These combined constraints are very similar to that coming from the PR3 release data,
\begin{equation}
\begin{split}
T_0 &= (2.89\pm 0.13) \:\rm{K} \\ % (\planck\:PR3)\\
	H_0 &=  (62.6\pm 3.8)  \: \rm{km \:s^{-1} Mpc^{-1}}.
\end{split}
\end{equation}

A common extension to \LCDM~is allowing for non-zero curvature $\Omega_K$. For this alone the ISW-lensing data does not add substantial additional information compared to the lensing spectrum. We noted though that Ref.~\cite{Bose:2020cjb}, using the \Planck~2015 likelihoods, and opening both $T_0$ and $\Omega_K$ (and combining with other data to break the large degeneracies), found a preference for a hotter and open Universe at high confidence, and we sought to test this result including $\hat C_\ell^{\phi T}$. However, irrespective of our ISW-lensing likelihood, we found that this preference completely disappears after updating their analysis from 2015 to 2018 PR3 \planck~data. This is because this preference was coupled to a very large optical depth\footnote{This was also speculated by the authors of Ref.~\cite{Bose:2020cjb} in private communication.}, which is excluded by the much tighter measurement of $\tau$ in the 2018 lowl.EE likelihood compared to 2015 (since usage of the High Frequency Instrument (HFI) data for this purpose was finally possible). This brings the preferred temperature and curvature of this analysis in good agreement with \LCDM.
\begin{table}%see chainloader.latextable
      \caption[]{68\% confidence regions on $H_0$ and $T_0$ from \planck~PR3 and PR4 data found in our \LCDM+$T_0$~MCMC chains with and without lensing and lensing-ISW data.}
         \label{table:T0H0}
         \begin{tabular}{|l|l|l|}
\hline
& $ H_0 $[$\rm{km}s^{-1}Mpc^{-1}$] & $ T_0$[$\rm{K}$] \\
\hline
\hline
PR3 CMB& $40^{+7}_{-10}$& $3.66^{+0.35}_{-0.23}$ \\
 CMB $+\hat C_L^{\phi \phi}$& $53^{+8}_{-6}$& $3.23^{+0.22}_{-0.29}$ \\
 CMB $+\hat C_L^{\phi T}$& $61.6\pm 4.4$& $2.92\pm 0.15$ \\
   CMB$+\hat C_L^{\phi \phi}+ \hat C_L^{\phi T}$& $62.6\pm 3.8$& $2.89\pm 0.13$ \\
   CMB +  $\hat C_L^{\phi \phi}$ ($T_0=T_{\rm FIRAS}$)& $67.40\pm 0.53$& $2.7255$ \\
\hline
\hline
PR4 CMB& $49^{+10}_{-9}$& $3.33^{+0.32}_{-0.35}$ \\
CMB  $+\hat C_L^{\phi \phi}$& $59.9^{+5.2}_{-4.0}$& $2.98^{+0.14}_{-0.18}$ \\
 CMB  $+\hat C_L^{\phi T}$& $62.9\pm 4.1$& $2.88\pm 0.14$ \\
 CMB  $+\hat C_L^{\phi \phi}+ \hat C_L^{\phi T}$& $63.5\pm 3.4$& $2.86\pm 0.12$ \\
   CMB   $+\hat C_L^{\phi \phi}$ ($T_0=T_{\rm FIRAS}$)& $67.23\pm 0.49$& $2.7255$ \\

\hline
 \end{tabular}
   \end{table}
\section{Conclusions}
With this paper we provide a new likelihood built from \planck~data that captures the lensing-ISW bispectrum information, by cross-correlating the \planck~lensing maps to the large-scale temperature. The signal, probing the low-redshift universe, is weak and detected at $4\sigma$ only. Current lensing spectrum data measurements are about 10 times more precise, so the new cross-spectrum band-powers are not expected to bring much new information in standard models. Nevertheless, these data points can prove useful in some extensions of \LCDM. Here we showed that they can successfully break the very strong degeneracy between the Hubble constant and the CMB temperature when constrained using CMB spectra alone. With the official 2018 \planck~release data~(PR3), the ISW-lensing constraint is almost twice as strong as that from the lensing spectrum. We also obtained results using the latest (and slightly more precise) CamSpec CMB likelihood and lensing results~\cite{Rosenberg:2022sdy, Carron:2022eyg}~(PR4). We found that combining all bandpowers gives very similar results for both releases, and consistency with the standard \LCDM~values. Of course the degeneracy remains strong, and our new \planck~internal joint measurement of the CMB temperature and Hubble constant do not come close to the precision in \LCDM~with fixed temperature. \JC{Nevertheless, the resulting Hubble constant best-fit value still lies in tension with local measurements by Ref.~\cite{Riess:2021jrx} by approximately $3\sigma$, with central value shifted even further from the local measurement value.} In models where the background evolution changes, external data, esp. baryon acoustic oscillation (BAO) data, can be a much more powerful way to break the geometric degeneracy. However, there may be extended models with background evolution consistent with BAO that modify late-time perturbation growth in such a way that the ISW likelihood still provides useful additional information.

The ISW-lensing likelihood \JC{is} available in two flavors at \url{https://github.com/carronj/planck_PR4_lensing}, and must be used in combination with the \planck~TT likelihoods in order to properly account for the covariance between these new data points and the existing \planck~public data.
\begin{acknowledgements}
JC acknowledges support from a SNSF Eccellenza Professorial Fellowship (No. 186879),
and AL support by the UK STFC grant ST/T000473/1. GF acknowledges the support of the European Research Council under the Marie Sk\l{}odowska Curie actions through the Individual Global Fellowship No.~892401 PiCOGAMBAS. JC thanks Lucas Lombriser and Benjamin Bose for useful discussions on $T_0$ and Ref.~\cite{Bose:2020cjb}. This research used resources of the National Energy Research Scientific Computing Center (NERSC), a U.S. Department of Energy Office of Science User Facility located at Lawrence Berkeley National Laboratory. Some of the results in this paper have been derived using the healpy and \HEALpix\ \footnote{\url{http://healpix.sourceforge.net}} packages~\cite{Zonca2019,Gorski:2004by}.
\end{acknowledgements}
%\onecolumngrid

\appendix
%\section{Covariances matrices and MCMC chains}

\begin{figure*}
   \centering
   \includegraphics[width=\hsize]{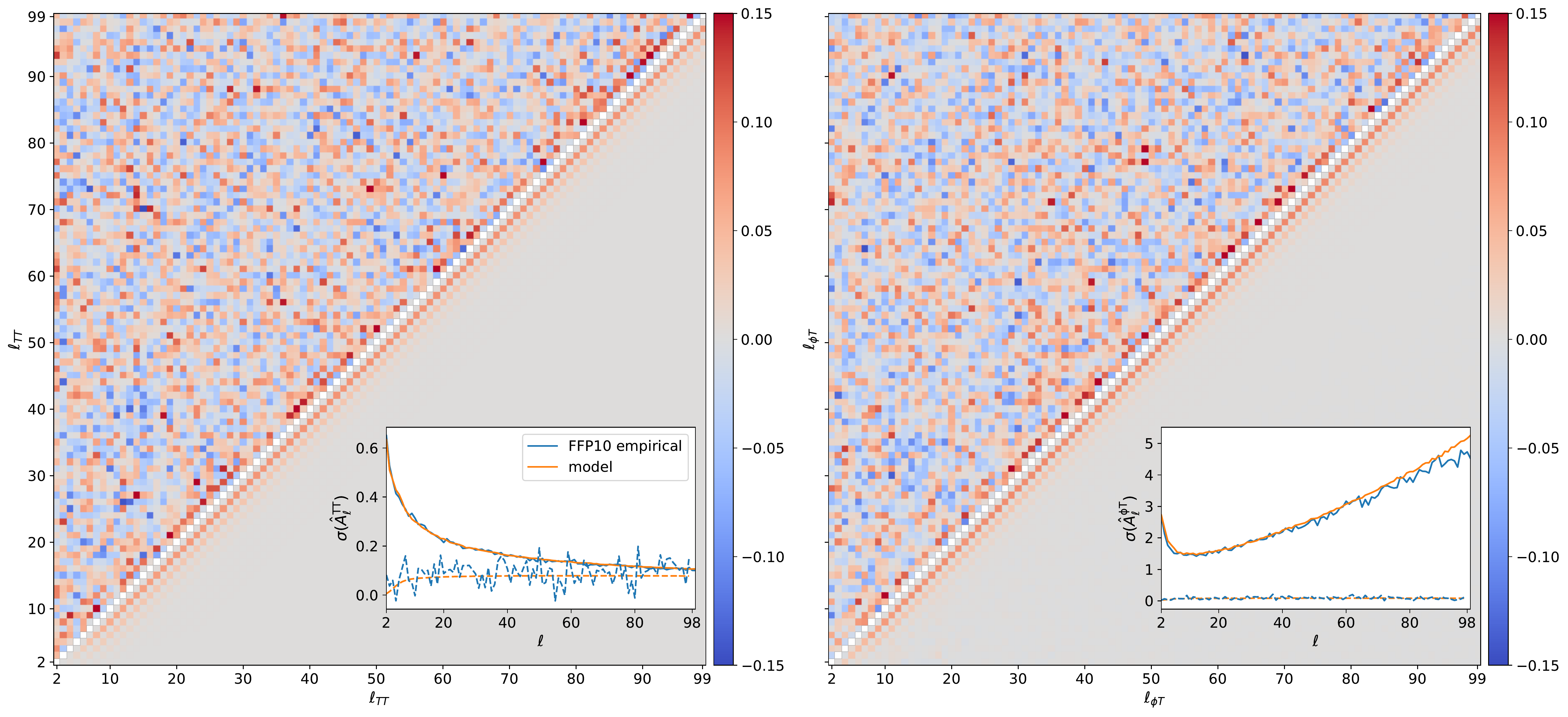}
   \caption{\label{fig:covs}Unbinned cross-correlation matrices for our unbinned $TT$ (left panel) and $\phi T$ (right panel) spectrum amplitude estimates, on the multipole range  $2\le \ell \le 99$, built on the lensing mask of $f_{\rm sky} = 67\%$. In each panel, the upper triangle shows the raw empirical covariance matrices obtained from the FFP10 simulation suite, and the lower triangle our refined model as described in the main text. The insets show the square root of the diagonal of the corresponding covariance matrices (solid lines, blue for the FFP10 empirical variances and orange for our model), together with the $\Delta_\ell = 2$ diagonal of the cross-correlation matrix, which is the most relevant offset diagonal owing to the approximate symmetry of the \planck~lensing mask with respect to the galactic equator.}
\end{figure*}
\begin{figure*}
   \centering
   \includegraphics[width=\hsize]{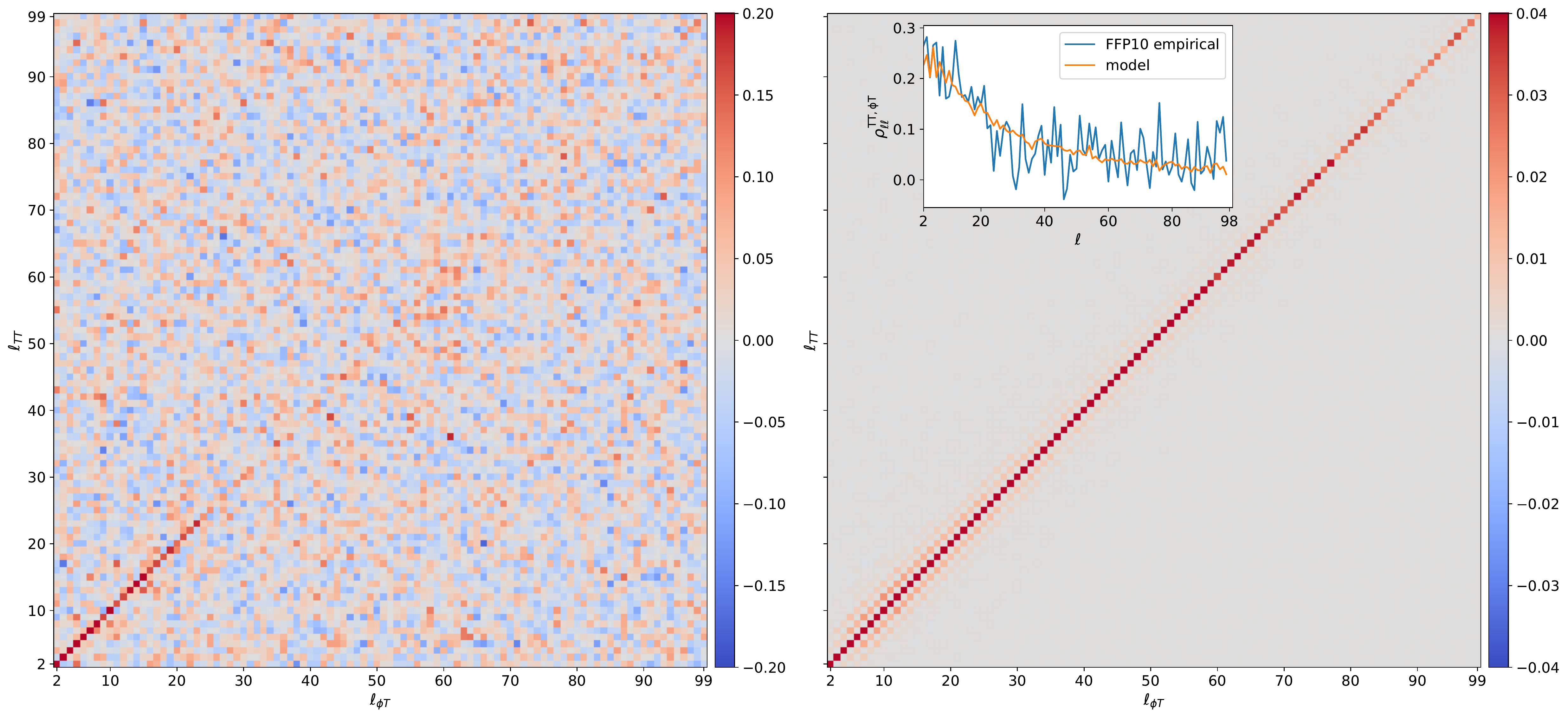}
   \caption{\label{fig:cov_TTPT}Unbinned cross-correlation matrix of the $TT$-$\phi T$ amplitude estimates on the multipole range  $2\le \ell_{\phi T}, \ell_{TT} \le 99$, built on the lensing mask of $f_{\rm sky} = 67\%$, as seen empirically on the FFP10 simulations (left panel) and for our refined model (right panel). The color scales are not identical. The inset show the diagonal elements for both cases (blue and orange respectively).}
\end{figure*}
\begin{figure*}
   \centering
   \includegraphics[width=\hsize]{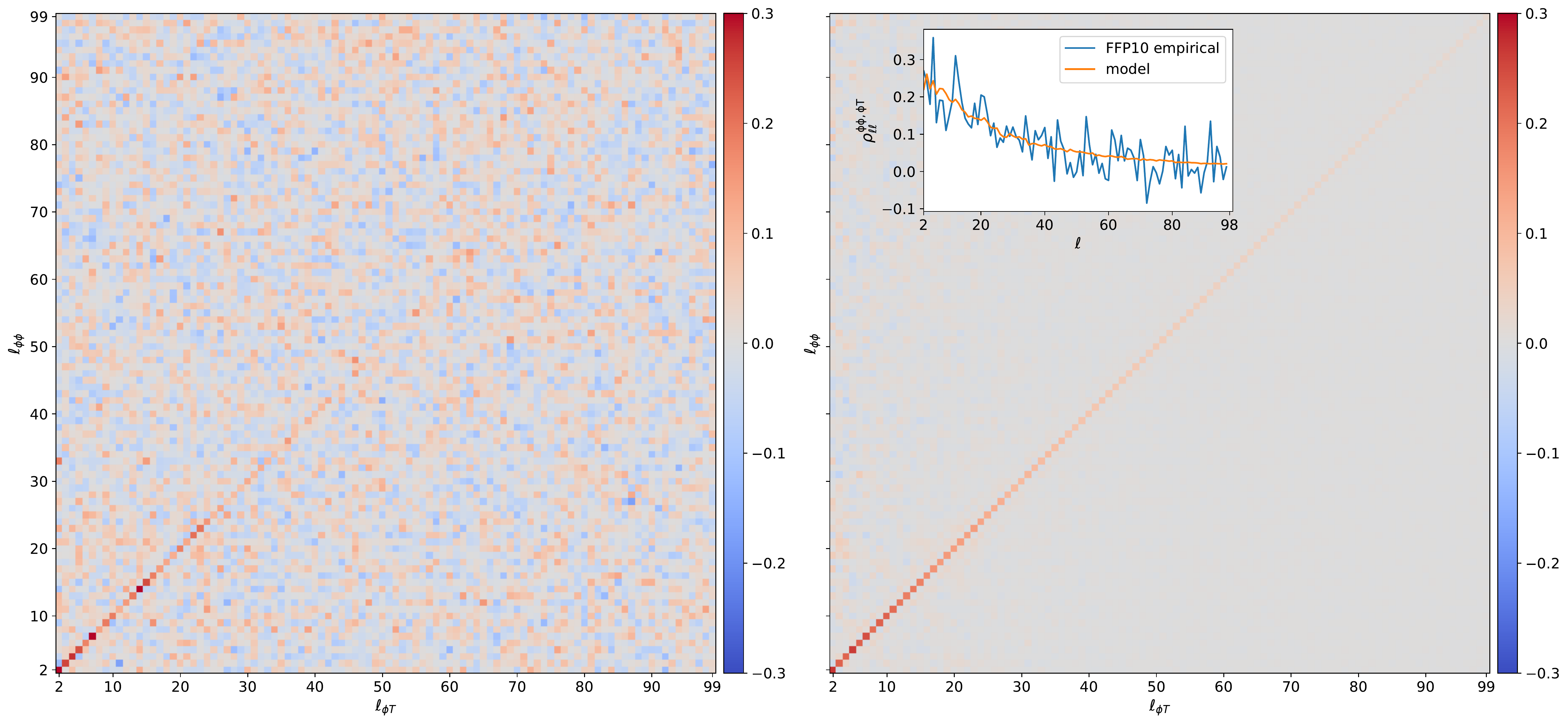}
   \caption{\label{fig:cov_PPTP}Unbinned cross-correlation matrix of the $T\phi$-$\phi \phi$ amplitude estimates, on the multipole range  $2\le \ell_{\phi T}, \ell_{\phi\phi} \le 99$, built on the lensing mask of $f_{\rm sky} = 67\%$, as seen empirically on the FFP10 simulations (left panel) and for our refined model (right panel) described in the text. The inset show the diagonal elements for both cases (blue and orange respectively).}
\end{figure*}

   \begin{figure*}
   \centering
   \includegraphics[width=\hsize]{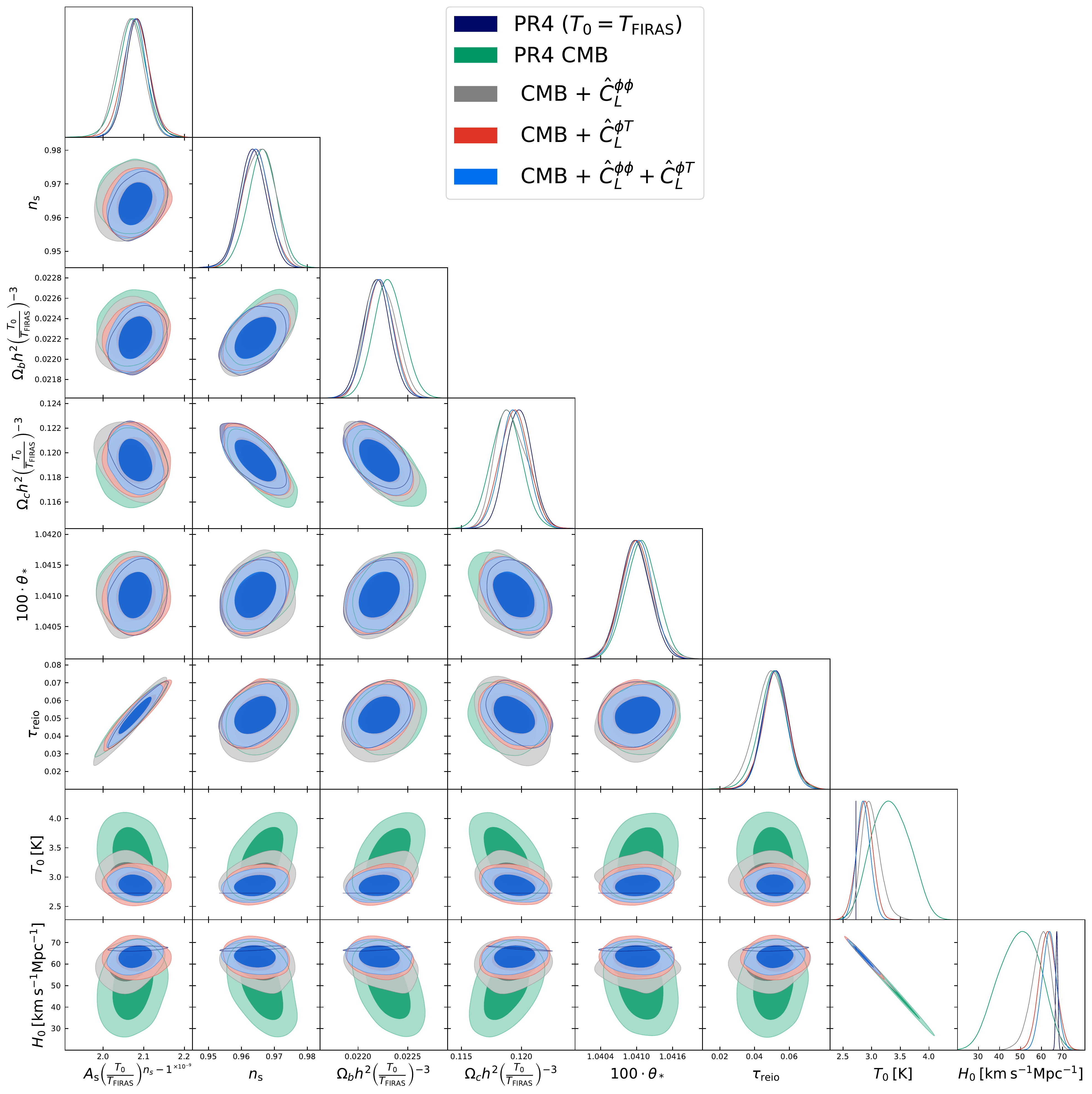}
      \caption{\label{fig:PR4mcmc}68\% pairwise confidence regions and marginal posteriors on the 7-parameters \LCDM+$T_0$ model for the \planck~PR4 (NPIPE) data, for CMB-only (green), CMB + lensing (grey), CMB + ISW-lensing (red), or all in combination (blue). Dark blue shows the PR4 \LCDM~only (CMB and lensing) results, where the CMB temperature is fixed to the FIRAS value. Fig.~\ref{fig:T0_mcmc} reproduces the $T_0$-$H_0$ subspace results and in comparison to PR3.
       }
  \end{figure*}
     \begin{figure*}
   \centering
   \includegraphics[width=\hsize]{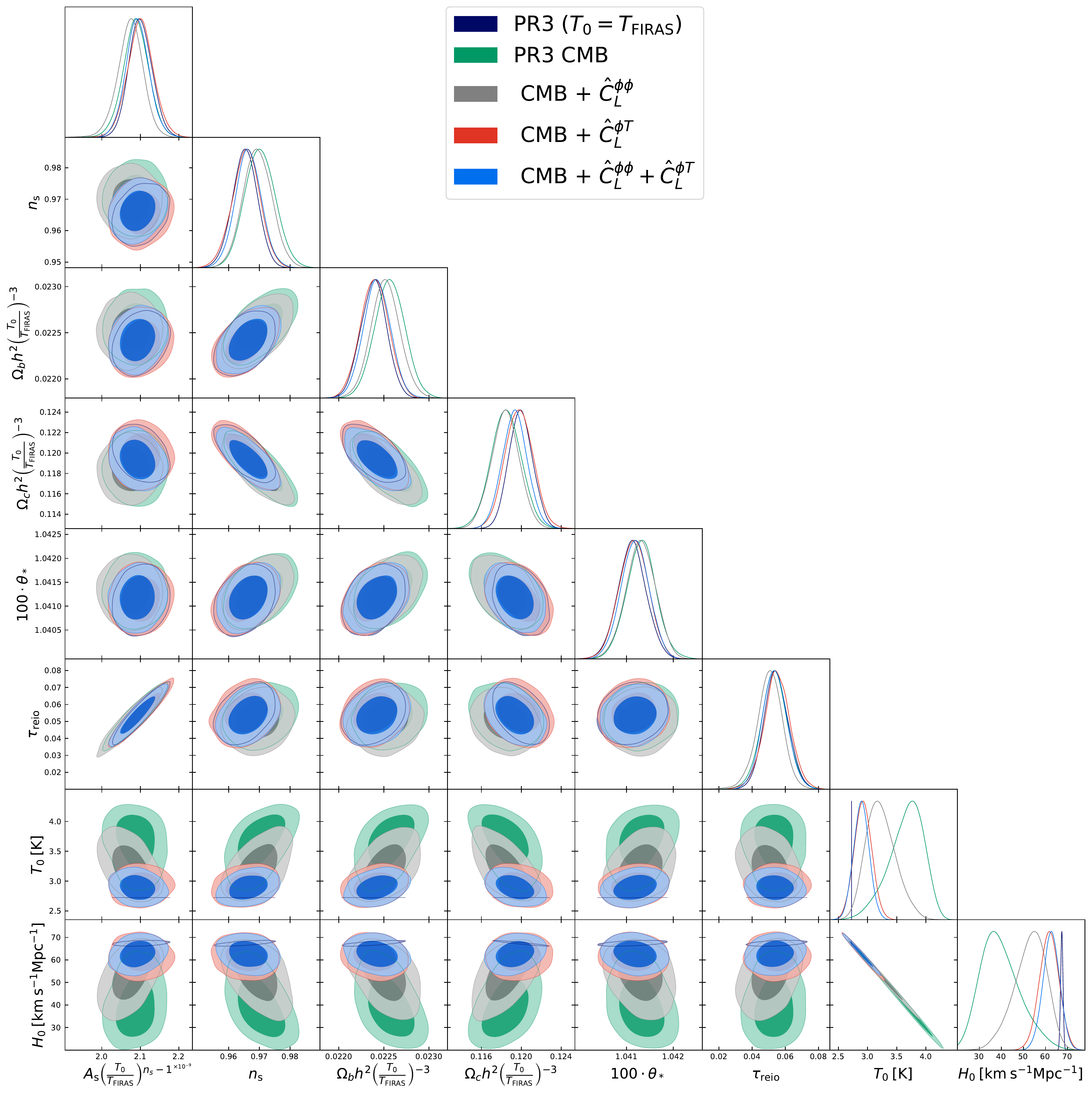}
      \caption{\label{fig:PR3mcmc} Same as Fig.~\ref{fig:PR4mcmc} for the \planck~2018 PR3 data}
  \end{figure*}

  \bibliography{texbase/cosmomc,texbase/antony,morebib}

\providecommand{\aj}{Astron. J. }\providecommand{\apj}{ApJ
  }\providecommand{\apjl}{ApJ
  }\providecommand{\mnras}{MNRAS}\providecommand{\prl}{PRL}\providecommand{\prd}{PRD}\providecommand{\jcap}{JCAP}\providecommand{\aap}{A\&A}
\begin{thebibliography}{39}%
\makeatletter
\providecommand \@ifxundefined [1]{%
 \@ifx{#1\undefined}
}%
\providecommand \@ifnum [1]{%
 \ifnum #1\expandafter \@firstoftwo
 \else \expandafter \@secondoftwo
 \fi
}%
\providecommand \@ifx [1]{%
 \ifx #1\expandafter \@firstoftwo
 \else \expandafter \@secondoftwo
 \fi
}%
\providecommand \natexlab [1]{#1}%
\providecommand \enquote  [1]{``#1''}%
\providecommand \bibnamefont  [1]{#1}%
\providecommand \bibfnamefont [1]{#1}%
\providecommand \citenamefont [1]{#1}%
\providecommand \href@noop [0]{\@secondoftwo}%
\providecommand \href [0]{\begingroup \@sanitize@url \@href}%
\providecommand \@href[1]{\@@startlink{#1}\@@href}%
\providecommand \@@href[1]{\endgroup#1\@@endlink}%
\providecommand \@sanitize@url [0]{\catcode `\\12\catcode `\$12\catcode
  `\&12\catcode `\#12\catcode `\^12\catcode `\_12\catcode `\%12\relax}%
\providecommand \@@startlink[1]{}%
\providecommand \@@endlink[0]{}%
\providecommand \url  [0]{\begingroup\@sanitize@url \@url }%
\providecommand \@url [1]{\endgroup\@href {#1}{\urlprefix }}%
\providecommand \urlprefix  [0]{URL }%
\providecommand \Eprint [0]{\href }%
\providecommand \doibase [0]{https://doi.org/}%
\providecommand \selectlanguage [0]{\@gobble}%
\providecommand \bibinfo  [0]{\@secondoftwo}%
\providecommand \bibfield  [0]{\@secondoftwo}%
\providecommand \translation [1]{[#1]}%
\providecommand \BibitemOpen [0]{}%
\providecommand \bibitemStop [0]{}%
\providecommand \bibitemNoStop [0]{.\EOS\space}%
\providecommand \EOS [0]{\spacefactor3000\relax}%
\providecommand \BibitemShut  [1]{\csname bibitem#1\endcsname}%
\let\auto@bib@innerbib\@empty
%</preamble>
\bibitem [{\citenamefont {Sachs}\ and\ \citenamefont {Wolfe}(1967)}]{Sachs:67}%
  \BibitemOpen
  \bibfield  {author} {\bibinfo {author} {\bibfnamefont {R.}~\bibnamefont
  {Sachs}}\ and\ \bibinfo {author} {\bibfnamefont {A.}~\bibnamefont {Wolfe}},\
  }\href@noop {} {\bibfield  {journal} {\bibinfo  {journal} {Astrophy. J.}\
  }\textbf {\bibinfo {volume} {147}},\ \bibinfo {pages} {735} (\bibinfo {year}
  {1967})}\BibitemShut {NoStop}%
\bibitem [{\citenamefont {Crittenden}\ and\ \citenamefont
  {Turok}(1996)}]{Crittenden:1995ak}%
  \BibitemOpen
  \bibfield  {author} {\bibinfo {author} {\bibfnamefont {R.~G.}\ \bibnamefont
  {Crittenden}}\ and\ \bibinfo {author} {\bibfnamefont {N.}~\bibnamefont
  {Turok}},\ }\bibfield  {title} {\bibinfo {title} {{Looking for $\Lambda$ with
  the Rees-Sciama Effect}},\ }\href
  {https://doi.org/10.1103/PhysRevLett.76.575} {\bibfield  {journal} {\bibinfo
  {journal} {\prd Lett.}\ }\textbf {\bibinfo {volume} {76}},\ \bibinfo {pages}
  {575} (\bibinfo {year} {1996})},\ \Eprint
  {https://arxiv.org/abs/astro-ph/9510072} {arXiv:astro-ph/9510072}
  \BibitemShut {NoStop}%
%%CITATION = ASTRO-PH/9510072;%%
\bibitem [{\citenamefont {Kable}\ \emph {et~al.}(2022)\citenamefont {Kable},
  \citenamefont {Benevento}, \citenamefont {Frusciante}, \citenamefont
  {De~Felice},\ and\ \citenamefont {Tsujikawa}}]{Kable:2021yws}%
  \BibitemOpen
  \bibfield  {author} {\bibinfo {author} {\bibfnamefont {J.~A.}\ \bibnamefont
  {Kable}}, \bibinfo {author} {\bibfnamefont {G.}~\bibnamefont {Benevento}},
  \bibinfo {author} {\bibfnamefont {N.}~\bibnamefont {Frusciante}}, \bibinfo
  {author} {\bibfnamefont {A.}~\bibnamefont {De~Felice}},\ and\ \bibinfo
  {author} {\bibfnamefont {S.}~\bibnamefont {Tsujikawa}},\ }\bibfield  {title}
  {\bibinfo {title} {{Probing modified gravity with integrated Sachs-Wolfe CMB
  and galaxy cross-correlations}},\ }\href
  {https://doi.org/10.1088/1475-7516/2022/09/002} {\bibfield  {journal}
  {\bibinfo  {journal} {JCAP}\ }\textbf {\bibinfo {volume} {09}},\ \bibinfo
  {pages} {002}},\ \Eprint {https://arxiv.org/abs/2111.10432} {arXiv:2111.10432
  [astro-ph.CO]} \BibitemShut {NoStop}%
\bibitem [{\citenamefont {Lesgourgues}\ \emph {et~al.}(2008)\citenamefont
  {Lesgourgues}, \citenamefont {Valkenburg},\ and\ \citenamefont
  {Gaztanaga}}]{Lesgourgues:2007ix}%
  \BibitemOpen
  \bibfield  {author} {\bibinfo {author} {\bibfnamefont {J.}~\bibnamefont
  {Lesgourgues}}, \bibinfo {author} {\bibfnamefont {W.}~\bibnamefont
  {Valkenburg}},\ and\ \bibinfo {author} {\bibfnamefont {E.}~\bibnamefont
  {Gaztanaga}},\ }\bibfield  {title} {\bibinfo {title} {{Constraining neutrino
  masses with the ISW-galaxy correlation function}},\ }\href
  {https://doi.org/10.1103/PhysRevD.77.063505} {\bibfield  {journal} {\bibinfo
  {journal} {Phys. Rev. D}\ }\textbf {\bibinfo {volume} {77}},\ \bibinfo
  {pages} {063505} (\bibinfo {year} {2008})},\ \Eprint
  {https://arxiv.org/abs/0710.5525} {arXiv:0710.5525 [astro-ph]} \BibitemShut
  {NoStop}%
\bibitem [{\citenamefont {Rees}\ and\ \citenamefont
  {Sciama}(1968)}]{ReesSciama}%
  \BibitemOpen
  \bibfield  {author} {\bibinfo {author} {\bibfnamefont {M.~J.}\ \bibnamefont
  {Rees}}\ and\ \bibinfo {author} {\bibfnamefont {D.~W.}\ \bibnamefont
  {Sciama}},\ }\bibfield  {title} {\bibinfo {title} {Large-scale density
  inhomogeneities in universe},\ }\href@noop {} {\bibfield  {journal} {\bibinfo
   {journal} {Nature}\ }\textbf {\bibinfo {volume} {217}},\ \bibinfo {pages}
  {511} (\bibinfo {year} {1968})}\BibitemShut {NoStop}%
\bibitem [{\citenamefont {Seljak}(1996)}]{SeljakISW}%
  \BibitemOpen
  \bibfield  {author} {\bibinfo {author} {\bibfnamefont {U.}~\bibnamefont
  {Seljak}},\ }\bibfield  {title} {\bibinfo {title} {{Rees-Sciama effect in a
  CDM universe}},\ }\href {https://doi.org/10.1086/176991} {\bibfield
  {journal} {\bibinfo  {journal} {Astrophys. J.}\ }\textbf {\bibinfo {volume}
  {460}},\ \bibinfo {pages} {549} (\bibinfo {year} {1996})},\ \Eprint
  {https://arxiv.org/abs/astro-ph/9506048} {arXiv:astro-ph/9506048}
  \BibitemShut {NoStop}%
\bibitem [{\citenamefont {Smith}\ \emph {et~al.}(2009)\citenamefont {Smith},
  \citenamefont {Hernandez-Monteagudo},\ and\ \citenamefont
  {Seljak}}]{SmithISW}%
  \BibitemOpen
  \bibfield  {author} {\bibinfo {author} {\bibfnamefont {R.~E.}\ \bibnamefont
  {Smith}}, \bibinfo {author} {\bibfnamefont {C.}~\bibnamefont
  {Hernandez-Monteagudo}},\ and\ \bibinfo {author} {\bibfnamefont
  {U.}~\bibnamefont {Seljak}},\ }\bibfield  {title} {\bibinfo {title} {{Impact
  of Scale Dependent Bias and Nonlinear Structure Growth on the ISW Effect:
  Angular Power Spectra}},\ }\href {https://doi.org/10.1103/PhysRevD.80.063528}
  {\bibfield  {journal} {\bibinfo  {journal} {Phys. Rev. D}\ }\textbf {\bibinfo
  {volume} {80}},\ \bibinfo {pages} {063528} (\bibinfo {year} {2009})},\
  \Eprint {https://arxiv.org/abs/0905.2408} {arXiv:0905.2408 [astro-ph.CO]}
  \BibitemShut {NoStop}%
\bibitem [{\citenamefont {Ferraro}\ \emph {et~al.}(2022)\citenamefont
  {Ferraro}, \citenamefont {Schaan},\ and\ \citenamefont
  {Pierpaoli}}]{Ferraro:2022twg}%
  \BibitemOpen
  \bibfield  {author} {\bibinfo {author} {\bibfnamefont {S.}~\bibnamefont
  {Ferraro}}, \bibinfo {author} {\bibfnamefont {E.}~\bibnamefont {Schaan}},\
  and\ \bibinfo {author} {\bibfnamefont {E.}~\bibnamefont {Pierpaoli}},\
  }\bibfield  {title} {\bibinfo {title} {{Is the Rees-Sciama effect detectable
  by the next generation of cosmological experiments?}},\ }\href@noop {} {\
  (\bibinfo {year} {2022})},\ \Eprint {https://arxiv.org/abs/2205.10332}
  {arXiv:2205.10332 [astro-ph.CO]} \BibitemShut {NoStop}%
\bibitem [{\citenamefont {Ade}\ \emph {et~al.}(2016{\natexlab{a}})\citenamefont
  {Ade} \emph {et~al.}}]{Planck:2015zfm}%
  \BibitemOpen
  \bibfield  {author} {\bibinfo {author} {\bibfnamefont {P.~A.~R.}\
  \bibnamefont {Ade}} \emph {et~al.} (\bibinfo {collaboration} {Planck}),\
  }\bibfield  {title} {\bibinfo {title} {{Planck 2015 results. XVII.
  Constraints on primordial non-Gaussianity}},\ }\href
  {https://doi.org/10.1051/0004-6361/201525836} {\bibfield  {journal} {\bibinfo
   {journal} {Astron. Astrophys.}\ }\textbf {\bibinfo {volume} {594}},\
  \bibinfo {pages} {A17} (\bibinfo {year} {2016}{\natexlab{a}})},\ \Eprint
  {https://arxiv.org/abs/1502.01592} {arXiv:1502.01592 [astro-ph.CO]}
  \BibitemShut {NoStop}%
\bibitem [{\citenamefont {{Planck Collaboration XV}}(2016)}]{Ade:2015zua}%
  \BibitemOpen
  \bibfield  {author} {\bibinfo {author} {\bibnamefont {{Planck Collaboration
  XV}}} (\bibinfo {collaboration} {Planck}),\ }\bibfield  {title} {\bibinfo
  {title} {{Planck 2015 results. XV. Gravitational lensing}},\ }\href
  {https://doi.org/10.1051/0004-6361/201525941} {\bibfield  {journal} {\bibinfo
   {journal} {\aap}\ }\textbf {\bibinfo {volume} {594}},\ \bibinfo {pages}
  {A15} (\bibinfo {year} {2016})},\ \Eprint {https://arxiv.org/abs/1502.01591}
  {arXiv:1502.01591 [astro-ph.CO]} \BibitemShut {NoStop}%
%%CITATION = ARXIV:1502.01591;%%
\bibitem [{\citenamefont {Boughn}\ and\ \citenamefont
  {Crittenden}(2004)}]{Boughn:2003yz}%
  \BibitemOpen
  \bibfield  {author} {\bibinfo {author} {\bibfnamefont {S.}~\bibnamefont
  {Boughn}}\ and\ \bibinfo {author} {\bibfnamefont {R.}~\bibnamefont
  {Crittenden}},\ }\bibfield  {title} {\bibinfo {title} {{A Correlation of the
  cosmic microwave sky with large scale structure}},\ }\href
  {https://doi.org/10.1038/nature02139} {\bibfield  {journal} {\bibinfo
  {journal} {Nature}\ }\textbf {\bibinfo {volume} {427}},\ \bibinfo {pages}
  {45} (\bibinfo {year} {2004})},\ \Eprint
  {https://arxiv.org/abs/astro-ph/0305001} {arXiv:astro-ph/0305001}
  \BibitemShut {NoStop}%
\bibitem [{\citenamefont {Ade}\ \emph {et~al.}(2016{\natexlab{b}})\citenamefont
  {Ade} \emph {et~al.}}]{Planck:2015fcm}%
  \BibitemOpen
  \bibfield  {author} {\bibinfo {author} {\bibfnamefont {P.~A.~R.}\
  \bibnamefont {Ade}} \emph {et~al.} (\bibinfo {collaboration} {Planck}),\
  }\bibfield  {title} {\bibinfo {title} {{Planck 2015 results. XXI. The
  integrated Sachs-Wolfe effect}},\ }\href
  {https://doi.org/10.1051/0004-6361/201525831} {\bibfield  {journal} {\bibinfo
   {journal} {Astron. Astrophys.}\ }\textbf {\bibinfo {volume} {594}},\
  \bibinfo {pages} {A21} (\bibinfo {year} {2016}{\natexlab{b}})},\ \Eprint
  {https://arxiv.org/abs/1502.01595} {arXiv:1502.01595 [astro-ph.CO]}
  \BibitemShut {NoStop}%
\bibitem [{\citenamefont {Aghanim}\ \emph
  {et~al.}(2020{\natexlab{a}})\citenamefont {Aghanim} \emph {et~al.}}]{PL2018}%
  \BibitemOpen
  \bibfield  {author} {\bibinfo {author} {\bibfnamefont {N.}~\bibnamefont
  {Aghanim}} \emph {et~al.} (\bibinfo {collaboration} {Planck}),\ }\bibfield
  {title} {\bibinfo {title} {{Planck 2018 results. VIII. Gravitational
  lensing}},\ }\href {https://doi.org/10.1051/0004-6361/201833886} {\bibfield
  {journal} {\bibinfo  {journal} {\aap}\ }\textbf {\bibinfo {volume} {641}},\
  \bibinfo {pages} {A8} (\bibinfo {year} {2020}{\natexlab{a}})},\ \Eprint
  {https://arxiv.org/abs/1807.06210} {arXiv:1807.06210 [astro-ph.CO]}
  \BibitemShut {NoStop}%
\bibitem [{\citenamefont {Carron}\ \emph {et~al.}(2022)\citenamefont {Carron},
  \citenamefont {Mirmelstein},\ and\ \citenamefont {Lewis}}]{Carron:2022eyg}%
  \BibitemOpen
  \bibfield  {author} {\bibinfo {author} {\bibfnamefont {J.}~\bibnamefont
  {Carron}}, \bibinfo {author} {\bibfnamefont {M.}~\bibnamefont
  {Mirmelstein}},\ and\ \bibinfo {author} {\bibfnamefont {A.}~\bibnamefont
  {Lewis}},\ }\bibfield  {title} {\bibinfo {title} {{CMB lensing from Planck
  PR4 maps}},\ }\href {https://doi.org/10.1088/1475-7516/2022/09/039}
  {\bibfield  {journal} {\bibinfo  {journal} {JCAP}\ }\textbf {\bibinfo
  {volume} {2022}}\bibfield  {number} {\bibinfo  {number} { (09)},\ \bibinfo
  {pages} {39}},\ }\Eprint {https://arxiv.org/abs/2206.07773} {arXiv:2206.07773
  [astro-ph.CO]} \BibitemShut {NoStop}%
\bibitem [{\citenamefont {Akrami}\ \emph {et~al.}(2020)\citenamefont {Akrami}
  \emph {et~al.}}]{Akrami:2020bpw}%
  \BibitemOpen
  \bibfield  {author} {\bibinfo {author} {\bibfnamefont {Y.}~\bibnamefont
  {Akrami}} \emph {et~al.} (\bibinfo {collaboration} {Planck}),\ }\bibfield
  {title} {\bibinfo {title} {{$Planck$ intermediate results. LVII. Joint Planck
  LFI and HFI data processing}},\ }\href
  {https://doi.org/10.1051/0004-6361/202038073} {\bibfield  {journal} {\bibinfo
   {journal} {Astron. Astrophys.}\ }\textbf {\bibinfo {volume} {643}},\
  \bibinfo {pages} {A42} (\bibinfo {year} {2020})},\ \Eprint
  {https://arxiv.org/abs/2007.04997} {arXiv:2007.04997 [astro-ph.CO]}
  \BibitemShut {NoStop}%
\bibitem [{\citenamefont {Fixsen}\ \emph {et~al.}(1996)\citenamefont {Fixsen},
  \citenamefont {Cheng}, \citenamefont {Gales}, \citenamefont {Mather},
  \citenamefont {Shafer},\ and\ \citenamefont {Wright}}]{Fixsen:1996nj}%
  \BibitemOpen
  \bibfield  {author} {\bibinfo {author} {\bibfnamefont {D.~J.}\ \bibnamefont
  {Fixsen}}, \bibinfo {author} {\bibfnamefont {E.~S.}\ \bibnamefont {Cheng}},
  \bibinfo {author} {\bibfnamefont {J.~M.}\ \bibnamefont {Gales}}, \bibinfo
  {author} {\bibfnamefont {J.~C.}\ \bibnamefont {Mather}}, \bibinfo {author}
  {\bibfnamefont {R.~A.}\ \bibnamefont {Shafer}},\ and\ \bibinfo {author}
  {\bibfnamefont {E.~L.}\ \bibnamefont {Wright}},\ }\bibfield  {title}
  {\bibinfo {title} {{The Cosmic Microwave Background spectrum from the full
  COBE FIRAS data set}},\ }\href {https://doi.org/10.1086/178173} {\bibfield
  {journal} {\bibinfo  {journal} {Astrophys. J.}\ }\textbf {\bibinfo {volume}
  {473}},\ \bibinfo {pages} {576} (\bibinfo {year} {1996})},\ \Eprint
  {https://arxiv.org/abs/astro-ph/9605054} {arXiv:astro-ph/9605054}
  \BibitemShut {NoStop}%
\bibitem [{\citenamefont {Schmittfull}\ \emph {et~al.}(2013)\citenamefont
  {Schmittfull}, \citenamefont {Challinor}, \citenamefont {Hanson},\ and\
  \citenamefont {Lewis}}]{Schmittfull:2013uea}%
  \BibitemOpen
  \bibfield  {author} {\bibinfo {author} {\bibfnamefont {M.~M.}\ \bibnamefont
  {Schmittfull}}, \bibinfo {author} {\bibfnamefont {A.}~\bibnamefont
  {Challinor}}, \bibinfo {author} {\bibfnamefont {D.}~\bibnamefont {Hanson}},\
  and\ \bibinfo {author} {\bibfnamefont {A.}~\bibnamefont {Lewis}},\ }\bibfield
   {title} {\bibinfo {title} {{On the joint analysis of CMB temperature and
  lensing-reconstruction power spectra}},\ }\href
  {https://doi.org/10.1103/PhysRevD.88.063012} {\bibfield  {journal} {\bibinfo
  {journal} {\prd}\ }\textbf {\bibinfo {volume} {88}},\ \bibinfo {pages}
  {063012} (\bibinfo {year} {2013})},\ \Eprint
  {https://arxiv.org/abs/1308.0286} {arXiv:1308.0286 [astro-ph.CO]}
  \BibitemShut {NoStop}%
%%CITATION = ARXIV:1308.0286;%%
\bibitem [{\citenamefont {Peloton}\ \emph {et~al.}(2017)\citenamefont
  {Peloton}, \citenamefont {Schmittfull}, \citenamefont {Lewis}, \citenamefont
  {Carron},\ and\ \citenamefont {Zahn}}]{Peloton:2016kbw}%
  \BibitemOpen
  \bibfield  {author} {\bibinfo {author} {\bibfnamefont {J.}~\bibnamefont
  {Peloton}}, \bibinfo {author} {\bibfnamefont {M.}~\bibnamefont
  {Schmittfull}}, \bibinfo {author} {\bibfnamefont {A.}~\bibnamefont {Lewis}},
  \bibinfo {author} {\bibfnamefont {J.}~\bibnamefont {Carron}},\ and\ \bibinfo
  {author} {\bibfnamefont {O.}~\bibnamefont {Zahn}},\ }\bibfield  {title}
  {\bibinfo {title} {{Full covariance of CMB and lensing reconstruction power
  spectra}},\ }\href {https://doi.org/10.1103/PhysRevD.95.043508} {\bibfield
  {journal} {\bibinfo  {journal} {\prd}\ }\textbf {\bibinfo {volume} {95}},\
  \bibinfo {pages} {043508} (\bibinfo {year} {2017})},\ \Eprint
  {https://arxiv.org/abs/1611.01446} {arXiv:1611.01446 [astro-ph.CO]}
  \BibitemShut {NoStop}%
%%CITATION = ARXIV:1611.01446;%%
\bibitem [{\citenamefont {{Frommert}}\ and\ \citenamefont
  {{En{\ss}lin}}(2009)}]{2009MNRAS.395.1837F}%
  \BibitemOpen
  \bibfield  {author} {\bibinfo {author} {\bibfnamefont {M.}~\bibnamefont
  {{Frommert}}}\ and\ \bibinfo {author} {\bibfnamefont {T.~A.}\ \bibnamefont
  {{En{\ss}lin}}},\ }\bibfield  {title} {\bibinfo {title} {{Ironing out
  primordial temperature fluctuations with polarization: optimal detection of
  cosmic structure imprints}},\ }\href
  {https://doi.org/10.1111/j.1365-2966.2009.14637.x} {\bibfield  {journal}
  {\bibinfo  {journal} {\mnras}\ }\textbf {\bibinfo {volume} {395}},\ \bibinfo
  {pages} {1837} (\bibinfo {year} {2009})}\BibitemShut {NoStop}%
\bibitem [{\citenamefont {Carron}(2013)}]{Carron:2012pw}%
  \BibitemOpen
  \bibfield  {author} {\bibinfo {author} {\bibfnamefont {J.}~\bibnamefont
  {Carron}},\ }\bibfield  {title} {\bibinfo {title} {{On the assumption of
  Gaussianity for cosmological two-point statistics and parameter dependent
  covariance matrices}},\ }\href {https://doi.org/10.1051/0004-6361/201220538}
  {\bibfield  {journal} {\bibinfo  {journal} {Astron. Astrophys.}\ }\textbf
  {\bibinfo {volume} {551}},\ \bibinfo {pages} {A88} (\bibinfo {year}
  {2013})},\ \Eprint {https://arxiv.org/abs/1204.4724} {arXiv:1204.4724
  [astro-ph.CO]} \BibitemShut {NoStop}%
\bibitem [{\citenamefont {Tegmark}(1997)}]{Tegmark:1996qt}%
  \BibitemOpen
  \bibfield  {author} {\bibinfo {author} {\bibfnamefont {M.}~\bibnamefont
  {Tegmark}},\ }\bibfield  {title} {\bibinfo {title} {How to measure cmb power
  spectra without losing information},\ }\href@noop {} {\bibfield  {journal}
  {\bibinfo  {journal} {\prd}\ }\textbf {\bibinfo {volume} {55}},\ \bibinfo
  {pages} {5895} (\bibinfo {year} {1997})},\ \Eprint
  {https://arxiv.org/abs/astro-ph/9611174} {astro-ph/9611174} \BibitemShut
  {NoStop}%
%%CITATION = ASTRO-PH/9611174;%%
\bibitem [{\citenamefont {Mirmelstein}\ \emph {et~al.}(2019)\citenamefont
  {Mirmelstein}, \citenamefont {Carron},\ and\ \citenamefont
  {Lewis}}]{Mirmelstein:2019sxi}%
  \BibitemOpen
  \bibfield  {author} {\bibinfo {author} {\bibfnamefont {M.}~\bibnamefont
  {Mirmelstein}}, \bibinfo {author} {\bibfnamefont {J.}~\bibnamefont
  {Carron}},\ and\ \bibinfo {author} {\bibfnamefont {A.}~\bibnamefont
  {Lewis}},\ }\bibfield  {title} {\bibinfo {title} {{Optimal filtering for CMB
  lensing reconstruction}},\ }\href
  {https://doi.org/10.1103/PhysRevD.100.123509} {\bibfield  {journal} {\bibinfo
   {journal} {Phys. Rev.}\ }\textbf {\bibinfo {volume} {D100}},\ \bibinfo
  {pages} {123509} (\bibinfo {year} {2019})},\ \Eprint
  {https://arxiv.org/abs/1909.02653} {arXiv:1909.02653 [astro-ph.CO]}
  \BibitemShut {NoStop}%
%%CITATION = ARXIV:1909.02653;%%
\bibitem [{\citenamefont {Lewis}\ \emph {et~al.}(2011)\citenamefont {Lewis},
  \citenamefont {Challinor},\ and\ \citenamefont {Hanson}}]{Lewis:2011fk}%
  \BibitemOpen
  \bibfield  {author} {\bibinfo {author} {\bibfnamefont {A.}~\bibnamefont
  {Lewis}}, \bibinfo {author} {\bibfnamefont {A.}~\bibnamefont {Challinor}},\
  and\ \bibinfo {author} {\bibfnamefont {D.}~\bibnamefont {Hanson}},\
  }\bibfield  {title} {\bibinfo {title} {{The shape of the CMB lensing
  bispectrum}},\ }\href {https://doi.org/10.1088/1475-7516/2011/03/018}
  {\bibfield  {journal} {\bibinfo  {journal} {\jcap}\ }\textbf {\bibinfo
  {volume} {1103}},\ \bibinfo {pages} {018} (\bibinfo {year} {2011})},\ \Eprint
  {https://arxiv.org/abs/1101.2234} {arXiv:1101.2234 [astro-ph.CO]}
  \BibitemShut {NoStop}%
%%CITATION = 1101.2234;%%
\bibitem [{\citenamefont {Hamimeche}\ and\ \citenamefont
  {Lewis}(2008)}]{Hamimeche:2008ai}%
  \BibitemOpen
  \bibfield  {author} {\bibinfo {author} {\bibfnamefont {S.}~\bibnamefont
  {Hamimeche}}\ and\ \bibinfo {author} {\bibfnamefont {A.}~\bibnamefont
  {Lewis}},\ }\bibfield  {title} {\bibinfo {title} {{Likelihood Analysis of CMB
  Temperature and Polarization Power Spectra}},\ }\href
  {https://doi.org/10.1103/PhysRevD.77.103013} {\bibfield  {journal} {\bibinfo
  {journal} {\prd}\ }\textbf {\bibinfo {volume} {77}},\ \bibinfo {pages}
  {103013} (\bibinfo {year} {2008})},\ \Eprint
  {https://arxiv.org/abs/0801.0554} {arXiv:0801.0554 [astro-ph]} \BibitemShut
  {NoStop}%
\bibitem [{\citenamefont {Fixsen}(2009)}]{Fixsen:2009ug}%
  \BibitemOpen
  \bibfield  {author} {\bibinfo {author} {\bibfnamefont {D.}~\bibnamefont
  {Fixsen}},\ }\bibfield  {title} {\bibinfo {title} {{The Temperature of the
  Cosmic Microwave Background}},\ }\href
  {https://doi.org/10.1088/0004-637X/707/2/916} {\bibfield  {journal} {\bibinfo
   {journal} {\apj}\ }\textbf {\bibinfo {volume} {707}},\ \bibinfo {pages}
  {916} (\bibinfo {year} {2009})},\ \Eprint {https://arxiv.org/abs/0911.1955}
  {arXiv:0911.1955 [astro-ph.CO]} \BibitemShut {NoStop}%
%%CITATION = ARXIV:0911.1955;%%
\bibitem [{\citenamefont {Gush}\ \emph {et~al.}(1990)\citenamefont {Gush},
  \citenamefont {Halpern},\ and\ \citenamefont {Wishnow}}]{PhysRevLett.65.537}%
  \BibitemOpen
  \bibfield  {author} {\bibinfo {author} {\bibfnamefont {H.~P.}\ \bibnamefont
  {Gush}}, \bibinfo {author} {\bibfnamefont {M.}~\bibnamefont {Halpern}},\ and\
  \bibinfo {author} {\bibfnamefont {E.~H.}\ \bibnamefont {Wishnow}},\
  }\bibfield  {title} {\bibinfo {title} {Rocket measurement of the
  cosmic-background-radiation mm-wave spectrum},\ }\href
  {https://doi.org/10.1103/PhysRevLett.65.537} {\bibfield  {journal} {\bibinfo
  {journal} {Phys. Rev. Lett.}\ }\textbf {\bibinfo {volume} {65}},\ \bibinfo
  {pages} {537} (\bibinfo {year} {1990})}\BibitemShut {NoStop}%
\bibitem [{\citenamefont {{Planck Collaboration XIII}}(2016)}]{Ade:2015xua}%
  \BibitemOpen
  \bibfield  {author} {\bibinfo {author} {\bibnamefont {{Planck Collaboration
  XIII}}} (\bibinfo {collaboration} {Planck}),\ }\bibfield  {title} {\bibinfo
  {title} {{Planck 2015 results. XIII. Cosmological parameters}},\ }\href
  {https://doi.org/10.1051/0004-6361/201525830} {\bibfield  {journal} {\bibinfo
   {journal} {\aap}\ }\textbf {\bibinfo {volume} {594}},\ \bibinfo {pages}
  {A13} (\bibinfo {year} {2016})},\ \Eprint {https://arxiv.org/abs/1502.01589}
  {arXiv:1502.01589 [astro-ph.CO]} \BibitemShut {NoStop}%
%%CITATION = ARXIV:1502.01589;%%
\bibitem [{\citenamefont {Ivanov}\ \emph {et~al.}(2020)\citenamefont {Ivanov},
  \citenamefont {Ali-Ha\"\i{}moud},\ and\ \citenamefont
  {Lesgourgues}}]{Ivanov:2020mfr}%
  \BibitemOpen
  \bibfield  {author} {\bibinfo {author} {\bibfnamefont {M.~M.}\ \bibnamefont
  {Ivanov}}, \bibinfo {author} {\bibfnamefont {Y.}~\bibnamefont
  {Ali-Ha\"\i{}moud}},\ and\ \bibinfo {author} {\bibfnamefont {J.}~\bibnamefont
  {Lesgourgues}},\ }\bibfield  {title} {\bibinfo {title} {{H0 tension or T0
  tension?}},\ }\href {https://doi.org/10.1103/PhysRevD.102.063515} {\bibfield
  {journal} {\bibinfo  {journal} {Phys. Rev. D}\ }\textbf {\bibinfo {volume}
  {102}},\ \bibinfo {pages} {063515} (\bibinfo {year} {2020})},\ \Eprint
  {https://arxiv.org/abs/2005.10656} {arXiv:2005.10656 [astro-ph.CO]}
  \BibitemShut {NoStop}%
\bibitem [{\citenamefont {Wen}\ \emph {et~al.}(2021)\citenamefont {Wen},
  \citenamefont {Scott}, \citenamefont {Sullivan},\ and\ \citenamefont
  {Zibin}}]{Wen:2020txi}%
  \BibitemOpen
  \bibfield  {author} {\bibinfo {author} {\bibfnamefont {Y.}~\bibnamefont
  {Wen}}, \bibinfo {author} {\bibfnamefont {D.}~\bibnamefont {Scott}}, \bibinfo
  {author} {\bibfnamefont {R.}~\bibnamefont {Sullivan}},\ and\ \bibinfo
  {author} {\bibfnamefont {J.~P.}\ \bibnamefont {Zibin}},\ }\bibfield  {title}
  {\bibinfo {title} {{Role of $T_0$ in CMB anisotropy measurements}},\ }\href
  {https://doi.org/10.1103/PhysRevD.104.043516} {\bibfield  {journal} {\bibinfo
   {journal} {Phys. Rev. D}\ }\textbf {\bibinfo {volume} {104}},\ \bibinfo
  {pages} {043516} (\bibinfo {year} {2021})},\ \Eprint
  {https://arxiv.org/abs/2011.09616} {arXiv:2011.09616 [astro-ph.CO]}
  \BibitemShut {NoStop}%
\bibitem [{\citenamefont {Aghanim}\ \emph
  {et~al.}(2020{\natexlab{b}})\citenamefont {Aghanim} \emph
  {et~al.}}]{PCP2018}%
  \BibitemOpen
  \bibfield  {author} {\bibinfo {author} {\bibfnamefont {N.}~\bibnamefont
  {Aghanim}} \emph {et~al.} (\bibinfo {collaboration} {Planck}),\ }\bibfield
  {title} {\bibinfo {title} {{Planck 2018 results. VI. Cosmological
  parameters}},\ }\href {https://doi.org/10.1051/0004-6361/201833910}
  {\bibfield  {journal} {\bibinfo  {journal} {\aap}\ }\textbf {\bibinfo
  {volume} {641}},\ \bibinfo {pages} {A6} (\bibinfo {year}
  {2020}{\natexlab{b}})},\ \Eprint {https://arxiv.org/abs/1807.06209}
  {arXiv:1807.06209 [astro-ph.CO]} \BibitemShut {NoStop}%
\bibitem [{\citenamefont {Lewis}\ \emph {et~al.}(2000)\citenamefont {Lewis},
  \citenamefont {Challinor},\ and\ \citenamefont {Lasenby}}]{Lewis:1999bs}%
  \BibitemOpen
  \bibfield  {author} {\bibinfo {author} {\bibfnamefont {A.}~\bibnamefont
  {Lewis}}, \bibinfo {author} {\bibfnamefont {A.}~\bibnamefont {Challinor}},\
  and\ \bibinfo {author} {\bibfnamefont {A.}~\bibnamefont {Lasenby}},\
  }\bibfield  {title} {\bibinfo {title} {{Efficient computation of CMB
  anisotropies in closed FRW models}},\ }\href {https://doi.org/10.1086/309179}
  {\bibfield  {journal} {\bibinfo  {journal} {\apj}\ }\textbf {\bibinfo
  {volume} {538}},\ \bibinfo {pages} {473} (\bibinfo {year} {2000})},\ \Eprint
  {https://arxiv.org/abs/astro-ph/9911177} {arXiv:astro-ph/9911177 [astro-ph]}
  \BibitemShut {NoStop}%
%%CITATION = ASTRO-PH/9911177;%%
\bibitem [{\citenamefont {Torrado}\ and\ \citenamefont
  {Lewis}(2021)}]{Torrado:2020dgo}%
  \BibitemOpen
  \bibfield  {author} {\bibinfo {author} {\bibfnamefont {J.}~\bibnamefont
  {Torrado}}\ and\ \bibinfo {author} {\bibfnamefont {A.}~\bibnamefont
  {Lewis}},\ }\bibfield  {title} {\bibinfo {title} {{Cobaya: Code for Bayesian
  Analysis of hierarchical physical models}},\ }\href
  {https://doi.org/10.1088/1475-7516/2021/05/057} {\bibfield  {journal}
  {\bibinfo  {journal} {\jcap}\ }\textbf {\bibinfo {volume} {05}},\ \bibinfo
  {pages} {057} (\bibinfo {year} {2021})},\ \Eprint
  {https://arxiv.org/abs/2005.05290} {arXiv:2005.05290 [astro-ph.IM]}
  \BibitemShut {NoStop}%
\bibitem [{\citenamefont {Seager}\ \emph {et~al.}(2000)\citenamefont {Seager},
  \citenamefont {Sasselov},\ and\ \citenamefont {Scott}}]{Seager:1999km}%
  \BibitemOpen
  \bibfield  {author} {\bibinfo {author} {\bibfnamefont {S.}~\bibnamefont
  {Seager}}, \bibinfo {author} {\bibfnamefont {D.~D.}\ \bibnamefont
  {Sasselov}},\ and\ \bibinfo {author} {\bibfnamefont {D.}~\bibnamefont
  {Scott}},\ }\bibfield  {title} {\bibinfo {title} {How exactly did the
  universe become neutral?},\ }\href@noop {} {\bibfield  {journal} {\bibinfo
  {journal} {\apjs}\ }\textbf {\bibinfo {volume} {128}},\ \bibinfo {pages}
  {407} (\bibinfo {year} {2000})},\ \Eprint
  {https://arxiv.org/abs/astro-ph/9912182} {astro-ph/9912182} \BibitemShut
  {NoStop}%
%%CITATION = ASTRO-PH 9912182;%%
\bibitem [{\citenamefont {Wong}\ \emph {et~al.}(2008)\citenamefont {Wong},
  \citenamefont {Moss},\ and\ \citenamefont {Scott}}]{Wong:2007ym}%
  \BibitemOpen
  \bibfield  {author} {\bibinfo {author} {\bibfnamefont {W.~Y.}\ \bibnamefont
  {Wong}}, \bibinfo {author} {\bibfnamefont {A.}~\bibnamefont {Moss}},\ and\
  \bibinfo {author} {\bibfnamefont {D.}~\bibnamefont {Scott}},\ }\bibfield
  {title} {\bibinfo {title} {{How well do we understand cosmological
  recombination?}},\ }\href@noop {} {\bibfield  {journal} {\bibinfo  {journal}
  {\mnras}\ }\textbf {\bibinfo {volume} {386}},\ \bibinfo {pages} {1023}
  (\bibinfo {year} {2008})},\ \Eprint {https://arxiv.org/abs/arXiv:0711.1357
  [astro-ph]} {arXiv:0711.1357 [astro-ph]} \BibitemShut {NoStop}%
%%CITATION = ARXIV:0711.1357;%%
\bibitem [{\citenamefont {Rosenberg}\ \emph {et~al.}(2022)\citenamefont
  {Rosenberg}, \citenamefont {Gratton},\ and\ \citenamefont
  {Efstathiou}}]{Rosenberg:2022sdy}%
  \BibitemOpen
  \bibfield  {author} {\bibinfo {author} {\bibfnamefont {E.}~\bibnamefont
  {Rosenberg}}, \bibinfo {author} {\bibfnamefont {S.}~\bibnamefont {Gratton}},\
  and\ \bibinfo {author} {\bibfnamefont {G.}~\bibnamefont {Efstathiou}},\
  }\bibfield  {title} {\bibinfo {title} {{CMB power spectra and cosmological
  parameters from Planck PR4 with CamSpec}},\ }\href@noop {} {\  (\bibinfo
  {year} {2022})},\ \Eprint {https://arxiv.org/abs/2205.10869}
  {arXiv:2205.10869 [astro-ph.CO]} \BibitemShut {NoStop}%
\bibitem [{\citenamefont {Bose}\ and\ \citenamefont
  {Lombriser}(2021)}]{Bose:2020cjb}%
  \BibitemOpen
  \bibfield  {author} {\bibinfo {author} {\bibfnamefont {B.}~\bibnamefont
  {Bose}}\ and\ \bibinfo {author} {\bibfnamefont {L.}~\bibnamefont
  {Lombriser}},\ }\bibfield  {title} {\bibinfo {title} {{Easing cosmic tensions
  with an open and hotter universe}},\ }\href
  {https://doi.org/10.1103/PhysRevD.103.L081304} {\bibfield  {journal}
  {\bibinfo  {journal} {Phys. Rev. D}\ }\textbf {\bibinfo {volume} {103}},\
  \bibinfo {pages} {L081304} (\bibinfo {year} {2021})},\ \Eprint
  {https://arxiv.org/abs/2006.16149} {arXiv:2006.16149 [astro-ph.CO]}
  \BibitemShut {NoStop}%
\bibitem [{\citenamefont {Riess}\ \emph {et~al.}(2022)\citenamefont {Riess}
  \emph {et~al.}}]{Riess:2021jrx}%
  \BibitemOpen
  \bibfield  {author} {\bibinfo {author} {\bibfnamefont {A.~G.}\ \bibnamefont
  {Riess}} \emph {et~al.},\ }\bibfield  {title} {\bibinfo {title} {{A
  Comprehensive Measurement of the Local Value of the Hubble Constant with 1 km
  $s^{-1}$ $\rm{Mpc}^{-1}$ Uncertainty from the Hubble Space Telescope and the
  SH0ES Team}},\ }\href {https://doi.org/10.3847/2041-8213/ac5c5b} {\bibfield
  {journal} {\bibinfo  {journal} {Astrophys. J. Lett.}\ }\textbf {\bibinfo
  {volume} {934}},\ \bibinfo {pages} {L7} (\bibinfo {year} {2022})},\ \Eprint
  {https://arxiv.org/abs/2112.04510} {arXiv:2112.04510 [astro-ph.CO]}
  \BibitemShut {NoStop}%
\bibitem [{\citenamefont {Zonca}\ \emph {et~al.}(2019)\citenamefont {Zonca},
  \citenamefont {Singer}, \citenamefont {Lenz}, \citenamefont {Reinecke},
  \citenamefont {Rosset}, \citenamefont {Hivon},\ and\ \citenamefont
  {Gorski}}]{Zonca2019}%
  \BibitemOpen
  \bibfield  {author} {\bibinfo {author} {\bibfnamefont {A.}~\bibnamefont
  {Zonca}}, \bibinfo {author} {\bibfnamefont {L.}~\bibnamefont {Singer}},
  \bibinfo {author} {\bibfnamefont {D.}~\bibnamefont {Lenz}}, \bibinfo {author}
  {\bibfnamefont {M.}~\bibnamefont {Reinecke}}, \bibinfo {author}
  {\bibfnamefont {C.}~\bibnamefont {Rosset}}, \bibinfo {author} {\bibfnamefont
  {E.}~\bibnamefont {Hivon}},\ and\ \bibinfo {author} {\bibfnamefont
  {K.}~\bibnamefont {Gorski}},\ }\bibfield  {title} {\bibinfo {title} {healpy:
  equal area pixelization and spherical harmonics transforms for data on the
  sphere in python},\ }\href {https://doi.org/10.21105/joss.01298} {\bibfield
  {journal} {\bibinfo  {journal} {Journal of Open Source Software}\ }\textbf
  {\bibinfo {volume} {4}},\ \bibinfo {pages} {1298} (\bibinfo {year}
  {2019})}\BibitemShut {NoStop}%
\bibitem [{\citenamefont {Gorski}\ \emph {et~al.}(2005)\citenamefont {Gorski}
  \emph {et~al.}}]{Gorski:2004by}%
  \BibitemOpen
  \bibfield  {author} {\bibinfo {author} {\bibfnamefont {K.~M.}\ \bibnamefont
  {Gorski}} \emph {et~al.},\ }\bibfield  {title} {\bibinfo {title} {Healpix --
  a framework for high resolution discretization, and fast analysis of data
  distributed on the sphere},\ }\href@noop {} {\bibfield  {journal} {\bibinfo
  {journal} {\apj}\ }\textbf {\bibinfo {volume} {622}},\ \bibinfo {pages} {759}
  (\bibinfo {year} {2005})},\ \Eprint {https://arxiv.org/abs/astro-ph/0409513}
  {astro-ph/0409513} \BibitemShut {NoStop}%
%%CITATION = ASTRO-PH 0409513;%%
\end{thebibliography}%
\end{document}